%% file: neurolog.tex
\documentclass[11pt,a4paper]{article}

\usepackage[utf8]{inputenc}
\usepackage[T1]{fontenc}
\usepackage{lmodern}
\usepackage[margin=1in]{geometry}
\usepackage{microtype}
\usepackage{authblk}
\usepackage{graphicx}
\usepackage{hyperref}
\usepackage{xcolor}
\usepackage{booktabs}
\usepackage{listings}
\usepackage{amsmath}
\usepackage{amssymb}
\usepackage{tikz}
\usetikzlibrary{positioning,shapes,arrows.meta,fit,backgrounds,calc,decorations.pathreplacing}
\usepackage[numbers,sort&compress]{natbib}
\usepackage{url}
\usepackage{enumitem}
\usepackage{xspace}
\usepackage{seqsplit}
\newcommand{\sys}{\textsc{NeuroLog}\xspace}

\definecolor{dlkw}{rgb}{0.0,0.0,0.6}
\definecolor{dlcom}{rgb}{0.3,0.5,0.3}
\definecolor{dlstr}{rgb}{0.6,0.0,0.0}
\lstdefinelanguage{datalog}{
  morekeywords={decl,input,output,Sym,Addr,Ver,Idx,number,unsigned,symbol},
  sensitive=true,
  morecomment=[l]{//},
  morestring=[b]",
}
\hypersetup{
  colorlinks=true,
  linkcolor=blue!50!black,
  citecolor=blue!50!black,
  urlcolor=blue!50!black,
}

\title{\textbf{\sys: Reasoning You Can Audit ---
Neuro-Symbolic Vulnerability Discovery via LLM Facts,
Datalog, and SMT}\\[6pt]
{\large\normalfont Technical Report}}

\author[1]{Sanjay Rawat}
\affil[1]{Independent Researcher, India}
\affil[ ]{\texttt{sanjayr@ymail.com}}

\date{}

\begin{document}

\maketitle

\input{sections/00_abstract}
\input{sections/01_introduction}
\input{sections/02_background}
\input{sections/03_motivation}
\input{sections/04_architecture}
\input{sections/05_methodology}
\input{sections/06_implementation}
\input{sections/07_evaluation}
\input{sections/08_discussion}
\input{sections/09_related}
\input{sections/10_conclusion}

\section*{Acknowledgements}
\noindent
We thank Anthropic Claude (Opus~4.7, 1M-context) for substantial
assistance in implementation, ablation, and writing. Decisions
about methodology, claims, and conclusions remain the sole
responsibility of the human author.

\bibliographystyle{plainnat}
\bibliography{refs}

\input{sections/A1_rules_appendix}

\end{document}

%% file: sections/00_abstract.tex
\begin{abstract}
\noindent
Vulnerability discovery on C and C++ source puts the analyst
between two awkward options. Heavyweight static analysers are
precise and auditable, but they want a working build before a
single query runs. Free-form LLMs read source without a build,
but on real codebases they invent details and lose track of
dataflow that crosses a function boundary. We present \sys, an
end-to-end \emph{compile-free} pipeline for source-level
vulnerability analysis that gives an LLM, Datalog, and SMT each
only the part of the job it is good at. An LLM reads one function
at a time and writes its dataflow as typed facts. A
Souffl\'e\xspace rule mesh composes those facts into
cross-function findings. A Z3 post-pass then drops the infeasible
findings and returns a satisfying model for each one that
survives. Where the source alone is not informative enough, we
fold in a small amount of runtime evidence: likely range
invariants, observed on a handful of corpus seeds, tighten the
SMT problem at near-zero extraction cost. A second LLM agent then
reads the model and writes a Python program that produces a
candidate crashing input, which an AddressSanitizer-instrumented
harness validates. The two combinations \sys composes are
established prior art: cheap static analysis narrowing SMT
(Saturn, Pinpoint), and Datalog with SMT (Formulog). What is new
here is that the fact base comes from a language model rather than
a compiler, that the whole pipeline runs without building the
target, and that the satisfying model is an artifact the crash
synthesiser consumes rather than a yes-or-no verdict.
Across stb, cJSON, a libxml2 scale run, an FFmpeg demuxer slice,
and two \texttt{curl} 8.3.0 audits, \sys re-discovers eight
published CVE-class issues end-to-end, among them the CVSS-9.8
\texttt{curl} SOCKS5 heap overflow (CVE-2023-38545), each one
confirmed by an AddressSanitizer harness. Beyond rediscovery, an
audit of libarchive HEAD surfaced five memory-safety bugs in
\emph{current} source, four of them previously unreported, across
the cpio reader and the XAR, WARC, and 7zip writers. All five were
filed upstream and several fixes have since merged, with a cpio
use-after-free regression acknowledged within seven hours.
Extraction is cheap, about 37\,s and \$0.005 of LLM cost on
\texttt{stb}. Crash synthesis turned a static finding into a
102-byte \texttt{stb\_vorbis} crash in two LLM iterations with no
fuzzer in the loop, and a likely-invariant filter built from three
Matroska seeds removes 13.2\% of the FFmpeg-demuxer feasible set,
including a static false positive the synthesiser had spent five
attempts trying to trigger.\footnote{A substantially revised
version of \sys, with an expanded evaluation and methodology, is
being prepared for a conference submission. Applying the current
system to recent releases of several widely-used C libraries has
surfaced more than twenty further previously-unreported
memory-safety issues, a number of them already fixed upstream (for
example in libyang). Those findings, and the extended design
behind them, will appear in that version.}
\end{abstract}

%% file: sections/01_introduction.tex
\section{Introduction}
\label{sec:intro}

Static vulnerability discovery on real C and C++ projects is a
long exercise in trade-offs. Industrial query engines such as
CodeQL~\cite{avgustinov2016ql} give reproducible findings, but
they require the target to be \emph{built}: an analyst chasing one
bug in one demuxer typically has to stand up the whole toolchain
first --- compilers, headers, flags, vendored dependencies,
sometimes a CI environment --- before a single query runs.
Code-property-graph tools~\cite{yamaguchi2014cpg} share that
auditability. Joern~\cite{joern} relaxes the build requirement,
since its C/C++ frontend parses directly and uses \texttt{gcc}
only for system headers, but constructing the graph still leaves
the analyst without a runtime- or SMT-grade decision procedure for
reachability. That compile-then-query cost is what keeps
declarative analysis out of reach for a quick, opportunistic look
at an unfamiliar target.

A second line of work asks whether an LLM can remove the build
step altogether. LLMs clearly help when paired with a classical
analyser~\cite{li2024iris}: they are good at inferring sources,
sinks, and sanitisers from documentation. Used free-form on a
whole codebase, though, they have two consistent weaknesses.
First, they confabulate, inventing function names, struct fields,
or call relations that read plausibly but are not in the source.
Second, they do not scale: a parser like FFmpeg~\cite{ffmpeg} has
hundreds of thousands of relevant facts, and even a million-token
window does not survive a plain ``read the codebase and find
bugs'' without the model losing track. (Very recent frontier
models handle this better, at a higher cost per target.) Recent
benchmark work~\cite{risse2025topscore} shows that detectors
trained at single-function granularity systematically over-claim
recall, since a function can be buggy only through the way its
caller invokes it. \sys's structural response is to keep the LLM
at per-function fact extraction and hand cross-function reasoning
to Datalog, where it can be audited.

This paper explores a different split. We use the LLM only where
it has a strong inductive bias: reading one C function and writing
its dataflow as typed Datalog facts, and reading a satisfying
model and emitting a Python program that produces a plausible
crashing input. Between those two moments we use formal reasoners
--- Datalog~\cite{jordan2016souffle} for cross-function
composition and Z3~\cite{demoura2008z3} for path-satisfiability.
The analyst need not compile the target: tree-sitter~\cite{tree-sitter}
parses incrementally, the LLM reads one function at a time, and
Souffl\'e\xspace runs on the fact base directly.

We call the system \sys. The two combinations it composes ---
cheap static analysis narrowing
SMT~\cite{xie2007saturn-toplas,shi2018pinpoint} and Datalog with
SMT~\cite{bembenek2020formulog} --- are established prior art
(\S\ref{sec:related}). What we contribute is the assembly: a
complete, compile-free vulnerability-analysis stack in which an
LLM, a Datalog rule mesh, and an SMT solver each handle the part
of the work the others are not suited to, with the satisfying
model itself flowing into a downstream LLM crash synthesiser.
Concretely:

\begin{itemize}[leftmargin=*]
\item \textbf{An LLM-as-fact-extractor design that removes the
build step.} Saturn, Pinpoint, SVF, CodeQL, and the Doop family
all require a working compilation of the target --- LLVM IR, Java
bytecode, or a build-wrapper invocation. We replace the
IR-lifting compiler with a function-by-function LLM that emits
typed Datalog facts (\texttt{Def}, \texttt{Use}, \texttt{ArithOp},
\texttt{Cast}, \texttt{Guard}, \texttt{FieldRead}, \texttt{Call}).
The LLM is constrained to the \emph{schema}: it does not invent
relations, it only fills tuples. That constraint is what keeps it
honest at scale (\S\ref{sec:methodology}).

\item \textbf{Datalog and SMT in a pipelined arrangement that
complements rather than competes with Formulog's intra-rule
combination~\cite{bembenek2020formulog}.} Souffl\'e\xspace runs to
fixpoint on the LLM-supplied facts and produces a finding catalog.
A separate Z3 pass then encodes, for each finding, only the
def-use chain plus the path guards leading to it (Phase~A).
Phase~B adds function summaries and depth-bounded callee inlining
for findings whose dataflow crosses a call. Souffl\'e\xspace stays
at full relational throughput, the SMT calls parallelise trivially
over findings (Phase~D), and the satisfying model is exposed to
downstream stages as an artifact rather than collapsed into a
yes-or-no verdict. An optional likely-invariant
pass~\cite{sahoo2013likely}, trained from a few corpus seeds,
tightens ranges further (Phase~E2), and every demotion is
preserved in a secondary tier rather than silently dropped
(\S\ref{sec:methodology}).

\item \textbf{Closed-loop crash synthesis from the SMT witness.}
For each SMT-feasible finding, a multi-shot LLM agent reads the
satisfying model, the chain of relevant facts, and a small
file-format hint, and writes a Python program that produces a
candidate crashing input. Candidates are validated against an
AddressSanitizer harness~\cite{asan}. On a no-crash, the
per-candidate ASan tail feeds back into the next round's prompt.
We calibrate this loop on \texttt{stb\_vorbis}
CVE-2023-45676~\cite{cve-2023-45676}: a 102-byte Ogg/Vorbis file
that crashes the harness deterministically emerges in two
iterations, with no fuzzer involved (\S\ref{sec:eval}). Saturn,
Pinpoint, and Snugglebug~\cite{chandra2009snugglebug} stop at
feasible-or-infeasible. The witness flowing back into a generator
is what \sys adds.

\item \textbf{An honest catalogue of where each layer works and
where it does not.} We re-discover eight CVE-class issues across
\texttt{stb\_vorbis}, cJSON~\cite{cjson}, FFmpeg, and curl (SOCKS
and WebSocket), including CVE-2023-38545 (CVSS 9.8), re-detected
through a new project-specific rule. We also surface five
memory-safety findings on libarchive~\cite{libarchive} HEAD, four
of them previously unreported, all filed upstream with several
fixes since merged. We document the misses too --- rule-mesh
coverage gaps, the static false positive that drove a
five-attempt synthesis miss, untriaged FFmpeg flags --- in
\S\ref{sec:eval} and \S\ref{sec:discussion}.
\end{itemize}

The rest of the paper is organised as follows.
\S\ref{sec:background} sketches the technologies the pipeline
composes. \S\ref{sec:motivation} walks through a single
\texttt{stb\_vorbis} bug end-to-end. That example carries the
narrative through the rest of the paper. \S\ref{sec:architecture}
describes the pipeline and \S\ref{sec:methodology} fills in the
phases. \S\ref{sec:implementation} reports implementation details
and \S\ref{sec:eval} presents the evaluation. \S\ref{sec:related}
positions \sys against neighbouring work and
\S\ref{sec:conclusion} closes.

%% file: sections/02_background.tex
\section{Background}
\label{sec:background}

\sys composes four bodies of work: declarative static analysis
with Datalog, query-based vulnerability hunting, SMT-driven
feasibility checking, and the recent literature on language models
for code. We give just enough background here to make the rest of
the paper self-contained. \S\ref{sec:related} returns to the same
literature with a comparative lens.

\paragraph{Datalog as a static-analysis IR.}
Andersen-style points-to analysis~\cite{andersen1994phd} can be
written in a few hundred lines of Horn clauses, and a generation
of declarative analyses --- Doop on Java~\cite{bravenboer2009doop},
\texttt{bddbddb} on context-sensitive
queries~\cite{whaley2004bddbddb}, IFDS-style interprocedural
distributive frameworks~\cite{reps1995ifds} --- has shown that
making the analysis declarative makes it auditable. Every claim of
the form ``variable $v$ at line $\ell$ is tainted'' is the
conclusion of a derivation tree whose leaves are concrete facts
from the program. Souffl\'e\xspace~\cite{jordan2016souffle}
synthesises C++ from Datalog rules and is the engine we use.
\sys's Datalog layer is roughly 30 rules organised in five passes:
alias, interprocedural taint, type-safety, memory-safety, and
sink. The rules are not this paper's main contribution. What is
new is what feeds them.

\paragraph{Code property graphs and CodeQL.}
The code-property-graph view~\cite{yamaguchi2014cpg} and
QL~\cite{avgustinov2016ql} both give the analyst a query language
over a unified representation of program structure. They share
Datalog's auditability while emphasising graph-reachability
patterns. Their cost is the build step: producing the IR (CodeQL
needs the project to compile under a special wrapper, and a CPG
needs Soot or a comparable IR producer for the host language).
\sys sits deliberately close to the Datalog school in its
reasoning layer and deliberately far from CodeQL on the front
end --- the LLM replaces the IR-lifting compiler.

\paragraph{Symbolic execution and SMT.}
Tools such as KLEE~\cite{cadar2008klee} and SAGE-class whitebox
fuzzers~\cite{godefroid2008automated} show that a SAT-style
back-end, usually Z3~\cite{demoura2008z3}, can take a program,
encode its execution along a path, and either find an input that
drives the program down that path or prove that none exists. The
same machinery works post-hoc, on a specific finding rather than a
whole-program execution: encode the def-use chain and the path
guards leading to a candidate bug site, then ask Z3 whether the bug
condition is reachable. This is what Phase~A and Phase~B of \sys
do, scoping the path explosion by bounding the chain. It is much
cheaper than a KLEE-style symbolic exploration, because the chain
is bounded by the Datalog finding rather than by the program's
branching factor.

\paragraph{LLMs for static analysis.}
Recent work --- IRIS~\cite{li2024iris} is the closest in spirit
--- shows that LLMs do well at the kinds of inference classical
static analysis is bad at: telling a sanitiser apart from a
wrapper, guessing which function an unfamiliar API name is a clone
of, reading documentation comments and inferring sources and
sinks. \sys leans on the same observation but moves the LLM
\emph{below} the rule mesh. Rather than taking Datalog's findings
and rewriting the rules, we let the LLM populate the \emph{facts}
the rules consume. Because the LLM is constrained to a typed
schema, and because Datalog ignores contradictory facts gracefully
(a redundant \texttt{Cast} tuple is an inefficiency, not a bug),
the LLM's mistakes show up as missing recall, not as fabricated
findings.

\paragraph{Likely invariants and runtime sanity.}
Likely invariants in the sense of
Sahoo~\textit{et al.}~\cite{sahoo2013likely} are program properties
observed to hold across a small number of normal-input runs --- a
counter never exceeds $7488$, an error code is always $0$. They
were originally proposed as a fault-localisation aid. We use them
as a precision filter on the SMT pass. The Daikon
project~\cite{ernst2007daikon} introduced a richer invariant
language for the same purpose. We deliberately restrict ourselves
to range invariants, because anything more expressive degrades the
``tier, don't drop'' guarantee we want over flagged findings.

\paragraph{Failure sketching.}
Kasikci~\textit{et al.}'s Gist system~\cite{kasikci2015failure}
introduced an adaptive backward-slice approach for root-causing
production failures: start with a small slice around the suspect
site, expand if necessary, and instrument only the locals the slice
mentions. We adapt the slice-doubling idea (\S\ref{sec:methodology})
for our LLM extraction budget. The analyst rarely wants to extract
facts for every function in a codebase at once, and a
doubled-radius slice gives a clean ``stop when growth stalls'' rule.

%% file: sections/03_motivation.tex
\section{A motivating example}
\label{sec:motivation}

We illustrate \sys on CVE-2023-45676 in
\texttt{stb\_vorbis}~\cite{stb,cve-2023-45676}. The bug has two
useful properties. The source fits on one slide and is obvious
once shown, yet it took GitHub Security Lab a manual review to
find, and it remains unfixed in upstream \texttt{master} despite
an advisory more than two years old. The interesting question is
not whether the pattern is detectable automatically, but which of
the obvious automation choices an analyst should reach for.

\paragraph{The bug.}
Listing~\ref{lst:vorbis-bug} shows the relevant region. The
function \texttt{start\_decoder} parses the Vorbis comment-list
header and allocates a string buffer per comment.

\begin{lstlisting}[language=C,basicstyle=\footnotesize\ttfamily,
                   numbers=left,numberstyle=\tiny\color{black!50},
                   firstnumber=3652,
                   escapeinside={(*@}{@*)},
                   xleftmargin=22pt,
                   float=tbp,
                   caption={\texttt{stb\_vorbis.c}, lines
                   3652--3676. \texttt{get32\_packet} reads
                   attacker-controlled bytes from the input
                   stream; \texttt{setup\_malloc} forwards to
                   \texttt{malloc}.  The product
                   \texttt{sizeof(char)\,*\,(len+1)} on line~3670
                   wraps to a small value when \texttt{len} is
                   close to \texttt{UINT32\_MAX}, the subsequent
                   loop on line~3673 then writes
                   \texttt{len}~bytes past the under-allocated
                   buffer.  ``\texttt{...}'' marks elided lines;
                   margin numbers track the true source.},
                   label=lst:vorbis-bug]
   len = get32_packet(f);                         /* taint source */
   f->vendor = (char*)setup_malloc(f, sizeof(char) * (len+1)); /* sink (alloc) */
   ...(*@\setcounter{lstnumber}{3659}@*)
   f->comment_list_length = get32_packet(f);      /* taint source */
   ...(*@\setcounter{lstnumber}{3663}@*)
   f->comment_list = (char**)setup_malloc(f, sizeof(char*) * f->comment_list_length); /* sink */
   ...(*@\setcounter{lstnumber}{3667}@*)
   for(i=0; i < f->comment_list_length; ++i) {
      len = get32_packet(f);                      /* taint source */
      f->comment_list[i] = (char*)setup_malloc(f, sizeof(char) * (len+1)); /* sink (alloc) */
      ...(*@\setcounter{lstnumber}{3672}@*)
      for(j=0; j < len; ++j) {                    /* OOB write    */
         f->comment_list[i][j] = get8_packet(f);
      }
   ...(*@\setcounter{lstnumber}{3676}@*)
   }
\end{lstlisting}

Two distinct failure modes sit in a 25-line region. Line~3664
multiplies an attacker-supplied 32-bit length by
\texttt{sizeof(char*)}, so a comment list of length $2^{29}$
wraps a 32-bit \texttt{size\_t} to zero. Line~3670 adds one to a
32-bit length, so \texttt{len = UINT32\_MAX} produces a zero-byte
allocation, and the loop on 3673 then writes \texttt{UINT32\_MAX}
bytes past it. Both rely on the same primitive: an
attacker-controlled \texttt{u32} reaches an arithmetic operation
whose result feeds \texttt{malloc}.

\paragraph{Why classical static analysis is awkward here.}
The pattern is within reach of CodeQL or any CPG-based query, but
both presuppose that the project has been built into an indexable
form. \texttt{stb} is a single-header library meant to be
\texttt{\#include}d into a host project, so running CodeQL on the
snippet means finding or writing a host that uses
\texttt{stb\_vorbis.c} as a \texttt{.c} file, building it under
the CodeQL wrapper, and only then writing the QL query and
iterating. Each step is easy on its own. The cumulative cost is
what rules out an opportunistic per-library check.

\paragraph{Why a free-form LLM is awkward here.}
A frontier LLM, asked to read \texttt{stb\_vorbis.c} (about 5,000
lines) and find ``integer-overflow bugs in allocation sizes,''
returns an answer in seconds. The answer is variably correct. In
our experiments the model would sometimes pick
\texttt{start\_decoder} immediately, sometimes pick a
syntactically similar but semantically uninteresting site, and
occasionally invent a bug at a line that does not exist. The
deeper problem is not the model's accuracy on a single prompt. It
is that the model's intuition does not compose: it gives no
derivation, no auditable evidence, and no way to ask ``would these
inputs really reach this site under the
\texttt{i < f->comment\_list\_length} guard?'' Packing more source
files into the prompt strains working memory by context length,
not by program structure.

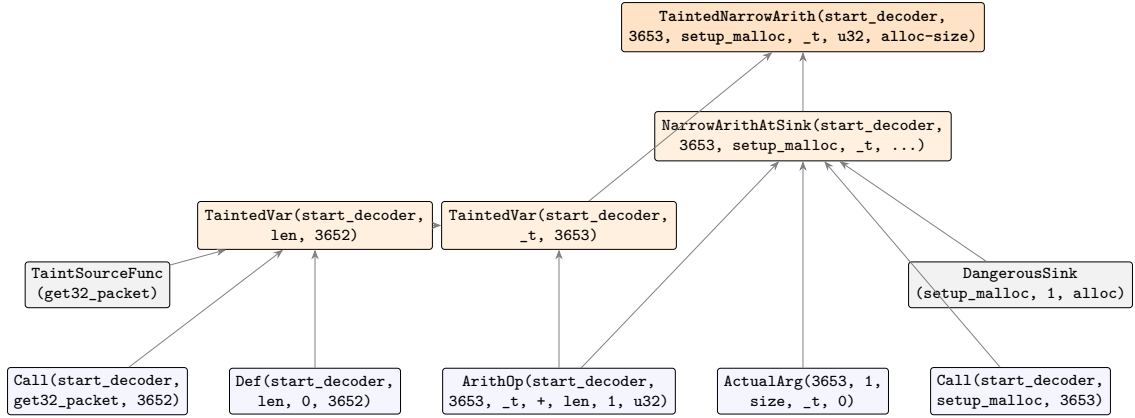
\begin{figure}[t]
\centering
\scalebox{0.78}{%
\begin{tikzpicture}[
  node distance=4mm and 7mm,
  font=\footnotesize,
  fact/.style={draw, rounded corners=1.5pt, fill=blue!4,
               inner sep=3pt, align=center, font=\scriptsize\ttfamily},
  cat/.style ={draw, rounded corners=1.5pt, fill=gray!10,
               inner sep=3pt, align=center, font=\scriptsize\ttfamily},
  derived/.style={draw, rounded corners=1.5pt, fill=orange!12,
               inner sep=3.5pt, align=center, font=\scriptsize\ttfamily},
  arrow/.style={-{Stealth[length=1.6mm]}, thin, gray}
]
  \node[fact]                 (c52)  {Call(start\_decoder,\\get32\_packet, 3652)};
  \node[fact, right=of c52]   (d52)  {Def(start\_decoder,\\len, 0, 3652)};
  \node[fact, right=of d52]   (a53)  {ArithOp(start\_decoder,\\3653, \_t, +, len, 1, u32)};
  \node[fact, right=of a53]   (aa53) {ActualArg(3653, 1,\\size, \_t, 0)};
  \node[fact, right=of aa53]  (c53)  {Call(start\_decoder,\\setup\_malloc, 3653)};

  \node[cat, above=10mm of c52]   (tsf) {TaintSourceFunc\\(get32\_packet)};
  \node[cat, above=10mm of c53]   (ds)  {DangerousSink\\(setup\_malloc, 1, alloc)};

  \node[derived, above=20mm of d52]  (tv1) {TaintedVar(start\_decoder,\\len, 3652)};
  \node[derived, above=20mm of a53]  (tv2) {TaintedVar(start\_decoder,\\\_t, 3653)};

  \node[derived, above=35mm of aa53] (nas) {NarrowArithAtSink(start\_decoder,\\3653, setup\_malloc, \_t, ...)};

  \node[derived, above=10mm of nas, fill=orange!22, font=\scriptsize\bfseries\ttfamily]
       (tna) {TaintedNarrowArith(start\_decoder,\\3653, setup\_malloc, \_t, u32, alloc-size)};

  \draw[arrow] (c52)  -- (tv1);
  \draw[arrow] (d52)  -- (tv1);
  \draw[arrow] (tsf)  -- (tv1);
  \draw[arrow] (tv1)  -- (tv2);
  \draw[arrow] (a53)  -- (tv2);
  \draw[arrow] (a53)  -- (nas);
  \draw[arrow] (aa53) -- (nas);
  \draw[arrow] (c53)  -- (nas);
  \draw[arrow] (ds)   -- (nas);
  \draw[arrow] (tv2)  -- (tna);
  \draw[arrow] (nas)  -- (tna);
\end{tikzpicture}}
\caption{Datalog derivation tree on the
\texttt{stb\_vorbis} CVE-2023-45676 region of
Listing~\ref{lst:vorbis-bug}.  Blue boxes are extracted facts
(LLM smell pass + tree-sitter); grey boxes are catalog facts
shipped with \sys; orange boxes are derived relations the rule
mesh produces.  The bold top tuple is the report the analyst
sees.  Every step in the derivation is mechanically auditable
and reproducible by re-running Souffl\'e\xspace on the same
fact base.}
\label{fig:deriv}
\end{figure}

\paragraph{What \sys does instead.}
\sys treats the LLM as a \emph{fact extractor} over one
function at a time. On \texttt{start\_decoder} it produces:
\begin{lstlisting}
  Def(start_decoder, len, 0, 3652)
  Call(start_decoder, get32_packet, 3652)
  ArithOp(start_decoder, 3653, _t, +, len, 1, "u32")
  Call(start_decoder, setup_malloc, 3653)
  ActualArg(3653, 1, size, _t, 0)
\end{lstlisting}
which a tree-sitter pre-pass complements with structural facts
an LLM would be wasteful for (CFG edges, dominance,
\texttt{FormalParam}, raw \texttt{Call} sites). Datalog rules
then compose the facts across the function into a derivation
that points at the bug (Figure~\ref{fig:deriv}).

\texttt{TaintSourceFunc(get32\_packet)} composes with the
extracted \texttt{Call}, \texttt{ActualArg}, and \texttt{Def}
facts to produce \texttt{TaintedVar(start\_decoder, len, 3652)}.
An \texttt{ArithOp} consuming that \texttt{TaintedVar} at the
\texttt{DangerousSink}-tagged \texttt{ActualArg} of
\texttt{setup\_malloc} yields \texttt{NarrowArithAtSink} at
line~3653, and the taint origin lifts it to the
\texttt{TaintedNarrowArith} that the report shows. None of these
rules know what \texttt{stb\_vorbis} is --- they are the same
rules we use on cJSON and FFmpeg.

\paragraph{From finding to crash.}
Phase~A of the symbolic-execution back-end encodes \texttt{len}
as a free 32-bit BitVec, walks the def-use chain back to its
\texttt{TaintSourceFunc} origin, and proves that the size argument
at \texttt{setup\_malloc} is attacker-controlled along a feasible
path. \sys hands that model, the def-use chain, and the local
snippet to a synthesis agent that knows Vorbis through Ogg --- the
\texttt{capture\_pattern} \texttt{0x4f676753},
\texttt{VORBIS\_packet\_comment}, and the rest. Its first witness
sets $\texttt{len} = \texttt{0xFFFFFFFF}$, but
\texttt{sizeof(char)\,*\,(len+1)} wraps to a zero-byte allocation
that does not fault. The ASan no-crash tail feeds back, and the
second witness sets $\texttt{len} = \texttt{0x7FFFFFFF}$, whose
size argument (computed in 64-bit \texttt{size\_t}) is a
\(\sim\)2\,GB request. The agent emits a Python program that
writes a 102-byte file. Fed to an ASan-instrumented harness, it
reports \texttt{allocation-size-too-big} at line~3653.

\paragraph{The total cost.}
On a 5-CPU laptop, tree-sitter scans the whole \texttt{stb}
codebase in under a second. The LLM extracts the $\sim$30
functions on the backward slice from \texttt{setup\_malloc} in
37\,s (Lite-tier model, parallel over functions). Souffl\'e\xspace
takes $\sim$2\,s and the SMT pass under a second, and synthesis
converges in two LLM iterations and $\sim$5\,min of wall time,
most of it network round-trips. End to end, the experience is
closer to a CTF write-up than to a CodeQL pipeline.

%% file: sections/04_architecture.tex
\section{Architecture}
\label{sec:architecture}

\begin{figure*}[t]
\centering
\scalebox{0.92}{%
\begin{tikzpicture}[
  node distance=4mm and 6mm,
  font=\footnotesize,
  stage/.style={draw, rounded corners=2pt, align=center, inner
                sep=4pt, minimum width=22mm, minimum height=10mm,
                fill=blue!4},
  llm/.style ={draw, rounded corners=2pt, align=center, inner
                sep=4pt, minimum width=22mm, minimum height=10mm,
                fill=orange!10},
  side/.style={draw, dashed, rounded corners=2pt, align=center,
                inner sep=3pt, minimum width=18mm, font=\scriptsize,
                fill=gray!5},
  arrow/.style={-{Stealth[length=2mm]}, thick}
]
  \node[stage] (src) {C source\\(no build)};
  \node[stage,right=of src] (slice) {1.\,Slice\\\scriptsize tree\_sitter};
  \node[llm,above=7mm of slice] (recon) {0.\,Recon\\\scriptsize LLM (opt.)};
  \node[llm,right=of slice]   (smell) {2.\,Smell pass\\\scriptsize LLM};
  \node[stage,right=of smell] (dl)    {3.\,Datalog\\\scriptsize Souffl\'e};
  \node[stage,right=of dl]    (smt)   {4.\,Symbex\\\scriptsize Z3};
  \node[llm,right=of smt]     (synth) {5.\,Synth\\\scriptsize LLM};
  \node[stage,above=7mm of synth] (asan)  {6.\,Harness\\\scriptsize ASan-instrumented};

  \draw[arrow] (recon) -- node[right,font=\scriptsize,xshift=-1pt]{seeds} (slice);
  \draw[arrow] (src)   -- (slice);
  \draw[arrow] (slice) -- (smell);
  \draw[arrow] (smell) -- (dl);
  \draw[arrow] (dl)    -- (smt);
  \draw[arrow] (smt)   -- (synth);
  \draw[arrow] (synth) -- (asan);

  \draw[arrow,bend left=55] (asan.east) to node[left,
       font=\scriptsize]{retry on no-crash} (synth.east);

  \node[side,below=8mm of slice] (e4) {E4 adaptive\\$\sigma$-double slice};
  \node[side,below=8mm of dl]    (e1) {E1 dependence\\filter};
  \node[side,below=8mm of smt]   (e2) {E2 likely\\range invariants};
  \node[side,below=8mm of synth] (e3) {E3 parser\\progress};

  \draw[->,dashed] (e4) -- (slice);
  \draw[->,dashed] (e1) -- (dl);
  \draw[->,dashed] (e2) -- (smt);
  \draw[->,dashed] (e3) -- (synth);
\end{tikzpicture}}
\caption{\sys's pipeline. Solid stages exchange artifacts
sequentially; orange stages are the two LLM moments (per-function
fact extraction, post-SMT crash synthesis); the dashed boxes are
the Phase E precision passes that attach to the relevant stage
boundaries. The arc from \emph{Harness} back to \emph{Synth}
is the Phase~C multi-shot retry loop. Crucially, no stage
requires the target to be compiled; the only built artifact is
the AddressSanitizer-instrumented validation harness in
stage~6, which the analyst supplies once per target. The optional
Phase~0 \emph{Recon} front-end (\S\ref{sec:methodology}) is a
cheap LLM read that proposes extra seed functions for the slice
and emits formally-checkable bug-class hypotheses consumed during
interpretation; it never emits findings itself.}
\label{fig:pipeline}
\end{figure*}
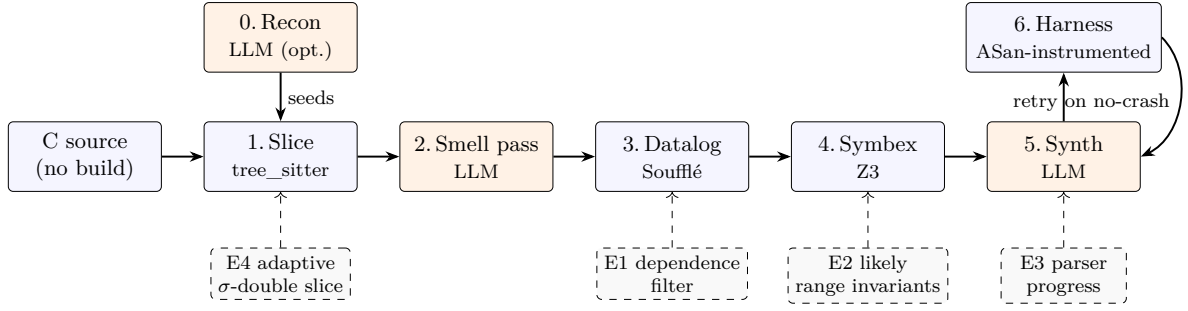

\sys is a six-stage pipeline. Each stage has one responsibility
and one output that the next stage consumes, so the analyst can
stop after any stage and keep a useful artifact: a slice, a fact
base, a Datalog finding catalog, an SMT-filtered catalog, or an
ASan-confirmed crashing input. \sys can run end-to-end on its own,
but the staging is there so that a user --- a security researcher
working from their own understanding of the codebase --- can drive
it interactively and inspect what each stage produced.
Figure~\ref{fig:pipeline} sketches the data flow, and
Table~\ref{tab:phases} maps each stage to its module, input,
output, and design role. An optional Phase~0 \emph{recon}
front-end can precede stage~1 to widen what the sink-seeded slice
would otherwise miss (\S\ref{sec:methodology}). It changes no
stage's contract.

\begin{table*}[t]
\centering
\footnotesize
\renewcommand{\arraystretch}{1.15}
\begin{tabular}{@{}llp{34mm}p{34mm}p{47mm}@{}}
\toprule
\# & Stage & Input & Output & Role \\
\midrule
1 & Slice
  & C source files
  & Backward slice (set of functions to extract from)
  & Tree-sitter parse + call-graph traversal from a curated
    sink catalog (\texttt{malloc}, \texttt{memcpy}, ffmpeg
    bitstream readers, etc.). No compilation, no IR. \\
2 & Smell pass
  & Per-function source + structural facts from tree-sitter
  & Function-local fact augmentation: \texttt{Def}/\texttt{Use},
    \texttt{ArithOp}, \texttt{Cast}, \texttt{Guard},
    \texttt{FieldRead}, \texttt{Call}, plus wrapper tags
  & A function-by-function LLM call. Constrained to the fact
    schema; the LLM cannot invent relations. \\
3 & Datalog
  & Reconciled fact base (mechanical floor + smell overlay)
  & Finding catalog (CSVs of \texttt{TaintedNarrowArith},
    \texttt{TaintedSizeAtSink}, \texttt{BufferOverflowInLoop}, \dots)
  & Five-pass Souffl\'e\xspace mesh: alias $\to$ interproc taint
    $\to$ type-safety $\to$ memory-safety $\to$ sink. \\
4 & Symbex
  & Each finding's \texttt{(func, addr, var, kind)} tuple
  & Per-finding verdict $\in$ \{feasible, infeasible, unknown\}
    + Z3 SAT model when feasible
  & Encodes the def-use chain plus path guards in Z3.
    Phase~A intra-procedural; Phase~B with function summaries +
    depth-bounded inlining. Phase~D parallelises across
    findings. \\
5 & Synth
  & SMT-feasible finding $+$ SAT model $+$ scaffold seed
  & Python emitter program $\to$ candidate input bytes
  & Multi-shot LLM agent. Reads the SAT model, def-use chain,
    file-format hint, and the scaffold; emits a Python program
    that produces $N$ candidate blobs per round. \\
6 & Harness
  & Candidate blobs
  & Crash transcript + parser-progress score
  & Runs each candidate through an
    AddressSanitizer-instrumented harness; on no crash, the
    failure transcript feeds the next Synth round. \\
\bottomrule
\end{tabular}
\caption{The six stages. Stages 2 and 5 are the two LLM moments;
3, 4, and 6 are formal/runtime; 1 is structural parsing. Phase E
(E1\,--\,E4) is auxiliary --- attaches to existing boundaries
without changing any stage's input/output contract.}
\label{tab:phases}
\end{table*}

\paragraph{Tier-don't-drop.}
Each precision step can demote a finding, but no step ever
silently drops one. When a Phase~E2 invariant demotes a finding
because its observed range excludes the bug condition, the finding
moves to a secondary tier with a back-pointer to the invariant
that demoted it. When E1 dependence filtering classifies a finding
as a downstream symptom of an upstream root cause, the back-pointer
leads to the root. This matters in practice, because the most
interesting findings --- attacker-controlled inputs the corpus
never exercised --- are exactly the ones a seed-driven invariant
pass would otherwise mute. We make the calibration visible in two
places. \S\ref{sec:methodology} documents how each pass exposes
its tier metadata, and \S\ref{sec:eval} reports the tier
breakdowns alongside the headline numbers, so a reader can re-tier
the data to their own tolerance.

\paragraph{Why six stages and not three.}
A simpler architecture would collapse stages 1 and 2 into ``the
LLM reads the codebase'' and stages 4, 5, and 6 into ``the LLM
produces crashing inputs.'' We report both ablations in
\S\ref{sec:eval}. The collapsed extractor loses the
structural-fact floor and confabulates. The collapsed synthesiser
loses the SAT-model anchor and produces blobs that are fluent but
rarely reach the bug site. The six-stage decomposition is what
keeps the LLM honest at the boundaries the rule mesh expects.

%% file: sections/05_methodology.tex
\section{Methodology}
\label{sec:methodology}

This section walks through each stage in detail, using the same
running example as \S\ref{sec:motivation}: the
\texttt{stb\_vorbis} comment-parsing region of
\texttt{start\_decoder} (CVE-2023-45676).

\subsection{Stage 0 --- recon (optional front-end)}
\label{sec:m-recon}

The backward slice (\S\ref{sec:m-slice}) is anchored on a catalog
of memory-safety sinks, so it never reaches functions that call no
such sink. That is precisely where a large and growing class of
bugs lives --- connection-reuse credential leaks, incomplete
validation, stale-state resets. The recon front-end closes that
gap without giving up the verifiability discipline. A cheap LLM
reads the target file and its callers and emits two things. First,
additional \emph{seed functions} that merge into the slice, so
policy, comparison, and cleanup functions enter the fact base.
Second, \emph{bug-class hypotheses}, each paired with a concrete
formal check: a Datalog rule sketch, an SMT constraint, or an ASan
trigger. Recon \emph{proposes, never judges}. It may raise a class
to investigate, but only a fired-or-silent formal rule resolves
one, and every hypothesis must carry a check or it is dropped.
Recon (in-pipeline, cheap model) and the host interpretation loop
that runs the checks communicate only through a persisted
artifact, so the two never need to be co-present.

The enabling primitive is a semantic fact the smell pass emits,
\texttt{FieldSemantic(func, field, role)}: ``this function plays
semantic \emph{role} (compared, validated, reset, and so on) over
this struct field.'' This is the neuro-symbolic bridge in
miniature. A syntactic extractor cannot know that
\texttt{Curl\_timestrcmp(a->user, b->user)} is a \emph{credential
comparison} without hardcoding every comparator name, but the LLM
resolves name to functionality and tags it
\texttt{role="compared"}. A small family of \emph{omission} rules
then proves, mechanically, that a handler exercising a role over
$\ge 2$ fields of a struct \emph{omits} a security-relevant field
the struct carries --- a reuse comparator that never compares a
credential, say, or a reset that leaves a secret stale. The rules
are self-scoping (the ``$\ge 2$ fields'' and ``two struct
instances'' tests identify comparators with no hardcoded function
list) and frame-driven (the struct's own fields, from a syntactic
struct scan, are the ground truth for what \emph{should} have been
handled). The LLM perceives. Datalog proves.

\subsection{Stage 1 --- slicing the function set}
\label{sec:m-slice}

\sys works on a backward slice from a small catalog of
\emph{sinks}: \texttt{malloc}-family allocators, \texttt{memcpy}
and friends, FFmpeg bitstream readers, and libc string functions.
Tree-sitter~\cite{tree-sitter} parses the codebase incrementally
and the call graph is built from identifier references inside
function bodies. The slice is the set of functions reachable
backward from any sink call site at depth $\sigma$.

\paragraph{Why backward instead of forward.}
A forward slice from the program's entry point is open-ended on
demuxer code where dispatch tables turn every public entry into
every function. Backward from sinks, the slice is bounded by what
Datalog will later query --- dataflow can only matter if it
touches a sink, and sinks are enumerable. The cost: forward-only
paths via function pointers the call graph does not resolve are
missed (\S\ref{sec:discussion}).

\paragraph{Adaptive depth (E4).}
The radius $\sigma$ trades extraction cost against recall on
deeply-nested wrappers. We use a $\sigma$-doubling
heuristic~\cite{kasikci2015failure}: compute the slice at
$\sigma\in\{2,4,8\}$ and take the smallest $\sigma$ at which the
growth-rate to the next falls under 5\,\%. On the FFmpeg demuxer
subset this recommends $\sigma{=}3$ (109/124/124 functions;
1.6\,\% growth at $\sigma{=}4$); on the small \texttt{stb}
codebase the heuristic does not converge and the maximum is
recommended. The output is a curation hint, not a hard filter.

\subsection{Stage 2 --- facts: mechanical floor + LLM smell pass}
\label{sec:m-facts}

The fact base is the pivot. Wrong facts cannot be recovered
downstream, and right facts let surprisingly modest Datalog rules
surface real bugs.

\paragraph{Mechanical floor.}
Tree-sitter contributes the structural facts an LLM would be
wasteful for: \texttt{FormalParam}, \texttt{Call} sites,
\texttt{CFGEdge}, dominance, basic-block heads. These are
deterministic and cheap, and the LLM never overrides them.

\paragraph{LLM smell pass.}
For each function in the slice the smell pass invokes a small
``Lite-tier'' LLM with the function source, line numbers, and an
instruction to emit a JSON list of facts using a fixed schema.
The schema is deliberately small (twelve relations) and
security-aware:

\begin{lstlisting}[basicstyle=\scriptsize\ttfamily]
Def(func, var, ver, addr)              ArithOp(func, addr, dst, op, src, operand, type)
Use(func, var, ver, addr)              Cast(func, addr, dst, src, kind, sw, dw, st, dt)
Call(func, callee, addr)               Guard(func, addr, var, op, rhs, polarity)
ActualArg(call_addr, idx, param, var)  FieldRead(func, addr, base, field)
ReturnVal(func, var, ver)              MemRead(func, addr, base, offset, size)
FormalParam(func, var, idx)            FieldWrite(func, addr, base, field, ...)
\end{lstlisting}

The prompt requires every emitted fact to be grounded in a
specific source line: a fabricated \texttt{ArithOp} pointing at a
line that contains no arithmetic must never appear. The fixed
schema is the corrective constraint --- the model cannot invent
non-existent relations, and when unsure it omits, which the rule
mesh handles gracefully (a missing \texttt{Cast} is a recall
miss, not a false positive).

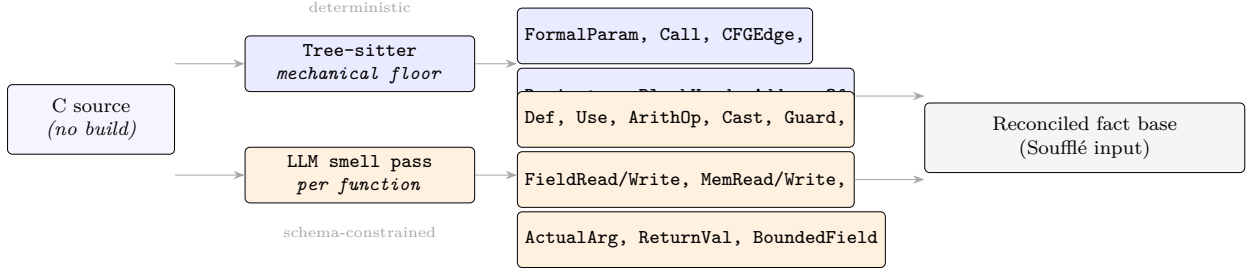
\begin{figure}[t]
\centering
\scalebox{0.92}{%
\begin{tikzpicture}[
  node distance=2mm and 6mm,
  font=\scriptsize,
  src/.style={draw, rounded corners=2pt, fill=blue!4,
              inner sep=4pt, align=center, minimum width=24mm,
              minimum height=10mm},
  prod/.style={draw, rounded corners=2pt, inner sep=3pt,
               align=center, minimum width=33mm, minimum height=8mm,
               font=\scriptsize\ttfamily},
  mech/.style={prod, fill=blue!8},
  llm/.style={prod, fill=orange!12},
  base/.style={draw, rounded corners=2pt, fill=gray!8,
               inner sep=4pt, align=center, minimum width=46mm,
               minimum height=10mm},
  arrow/.style={-{Stealth[length=1.6mm]}, thin, gray!70}
]
  \node[src]                       (csrc) {C source\\ \textit{(no build)}};
  \node[mech, right=10mm of csrc, yshift=8mm]
                                   (mech) {Tree-sitter\\\textit{mechanical floor}};
  \node[llm,  right=10mm of csrc, yshift=-8mm]
                                   (llm)  {LLM smell pass\\\textit{per function}};

  \node[mech, right=of mech, yshift=4mm,  minimum width=40mm]
                                   (m1)   {FormalParam, Call, CFGEdge,};
  \node[mech, below=0.5mm of m1.south west, anchor=north west, minimum width=40mm]
                                   (m2)   {Dominates, BlockHead, AddressOf};

  \node[llm,  right=of llm, yshift=8mm, minimum width=40mm]
                                   (l1)   {Def, Use, ArithOp, Cast, Guard,};
  \node[llm,  below=0.5mm of l1.south west, anchor=north west, minimum width=40mm]
                                   (l2)   {FieldRead/Write, MemRead/Write,};
  \node[llm,  below=0.5mm of l2.south west, anchor=north west, minimum width=40mm]
                                   (l3)   {ActualArg, ReturnVal, BoundedField};

  \node[base, right=10mm of $(m2.east)!0.5!(l2.east)$]
                                   (fb)   {Reconciled fact base\\(Souffl\'e\xspace input)};

  \draw[arrow] (csrc.east|-mech.west)  -- (mech.west);
  \draw[arrow] (csrc.east|-llm.west)   -- (llm.west);
  \draw[arrow] (mech.east) -- (m1.west|-mech.east);
  \draw[arrow] (llm.east)  -- (l1.west|-llm.east);
  \draw[arrow] (m2.east)   -- (fb.west|-m2.east);
  \draw[arrow] (l2.east)   -- (fb.west|-l2.east);

  \node[font=\tiny, gray!70, above=2mm of mech] {deterministic};
  \node[font=\tiny, gray!70, below=2mm of llm]  {schema-constrained};
\end{tikzpicture}}
\caption{Producer split for the twelve-relation fact schema.
The mechanical floor (blue) handles relations a syntactic
traversal can decide; the LLM smell pass (orange) handles
semantic ones --- which assignment is a \texttt{Def}, which
conditional is a \texttt{Guard} on a tainted variable, which
access is a \texttt{FieldRead}. Both streams reconcile into a
single Souffl\'e\xspace input. The LLM is constrained to the
schema: it can only fill tuples, never invent relation names.}
\label{fig:extractor-split}
\end{figure}

Figure~\ref{fig:extractor-split} maps each relation to its
producer: mechanical floor for structural facts, LLM smell pass
for semantic ones a syntactic AST traversal cannot decide.

\paragraph{Concretely.}
On \texttt{start\_decoder} the smell pass emits, among many
others, the facts shown in Listing~\ref{lst:vorbis-facts}.

\begin{lstlisting}[basicstyle=\scriptsize\ttfamily,
    float=tbp,
    caption={Subset of facts the smell pass emits on the
    \texttt{stb\_vorbis} region of Listing~\ref{lst:vorbis-bug},
    formatted as Souffl\'e\xspace input rows.},
    label=lst:vorbis-facts]
Call(start_decoder, get32_packet, 3652)
Def(start_decoder, len, 0, 3652)
Use(start_decoder, len, 0, 3653)
ArithOp(start_decoder, 3653, _t, +, len, 1, "u32")
Call(start_decoder, setup_malloc, 3653)
ActualArg(3653, 1, size, _t, 0)
Call(start_decoder, get32_packet, 3669)
Def(start_decoder, len, 0, 3669)
ArithOp(start_decoder, 3670, _t2, +, len, 1, "u32")
Call(start_decoder, setup_malloc, 3670)
ActualArg(3670, 1, size, _t2, 0)
\end{lstlisting}

The mechanical pass adds \texttt{CFGEdge}, \texttt{FormalParam},
and \texttt{Call} sites for any helpers reachable from this
function.

\paragraph{Cost vs.\ recall.}
On the \texttt{stb} slice (128 functions, 18.6 KLoC) the smell
pass takes 36.6\,s at \$0.005 with a Lite-tier model and 15-way
parallelism. The same slice with a frontier model extracting
facts alone (no mechanical floor) takes 55\,min at \$0.30--\$0.50 ---
$90{\times}$ slower, $70{\times}$ more expensive, and on
bug-relevant functions \emph{equal recall} (\S\ref{sec:eval}).

\subsection{Stage 3 --- Datalog rule mesh}
\label{sec:m-datalog}

The Datalog layer composes facts across functions in five passes
--- (i) alias, (ii) interprocedural taint, (iii) type safety,
(iv) memory safety, (v) sink --- with Souffl\'e\xspace's
stratified negation making the pass ordering explicit. The mesh
is $\sim$30 rules covering tainted size arguments, allocation
overflow, sign-confusion casts, narrowing-cast at sinks,
buffer-overflow in loops, double-free, use-after-free,
NULL-deref of unchecked allocation, and unbounded
counter-as-index. Each family is one or two relations composing
dataflow facts in a single rule (Appendix~\ref{app:rules}).

\paragraph{Worked example.}
The CVE-2023-45676 finding emerges as the conjunction of four
relations.  First, \texttt{TaintSourceFunc(get32\_packet)} is in
the catalog. Source taint propagates to \texttt{TaintedVar}:

\begin{lstlisting}[language=datalog,basicstyle=\scriptsize\ttfamily]
TaintedVar(f, v, addr, "ret_from_get32_packet", "ret") :-
    Call(f, "get32_packet", addr),
    Def(f, v, _, addr).
\end{lstlisting}

which fires on the \texttt{Def(start\_decoder, len, 0, 3652)}
above.  Standard taint propagation through arithmetic gives
\texttt{TaintedVar(start\_decoder, \_t, 3653, \dots)} for the
result of \texttt{len + 1}. The \texttt{NarrowArithAtSink}
relation looks for arithmetic in a declared-narrow type that flows
into a sink argument; combined with \texttt{TaintedVar}, it
produces:

\begin{lstlisting}[language=datalog,basicstyle=\scriptsize\ttfamily]
TaintedNarrowArith(f, ca, callee, dst, w, sign, risk, origin) :-
    NarrowArithAtSink(f, ca, callee, dst, arith_a, w, sign, risk),
    TaintedVar(f, dst, arith_a, origin, _).
\end{lstlisting}

That tuple is the report the analyst sees:
\texttt{start\_decoder @ 3653, var=\_t, narrow=u32,
risk=alloc-size}.

\paragraph{What the Datalog layer is \emph{not} doing.}
It is not deciding feasibility. The taint propagation is a
\emph{may}-flow over-approximation. Only the SMT pass below
decides whether a path is satisfiable. The split is by design:
Datalog stays at full relational throughput, with no SAT calls in
rule bodies, in contrast to Formulog~\cite{bembenek2020formulog},
and SMT stays per-finding and embarrassingly parallel
(\S\ref{sec:m-symbex}).

\paragraph{Closing an LLM extraction gap (Pass 6).}
On the libxml2 case study (\S\ref{sec:eval-cve}) the agent
identified a recurring extraction artifact: for an inlined size
expression like \verb|xmlMalloc(lenn + lenp + 2)|, the smell pass
emits an \texttt{ArithOp} with destination \texttt{tmp} (a
synthetic name) but no \texttt{Def} or \texttt{VarType} on it,
while \texttt{ActualArg} records the call's argument as the
literal expression string \verb|"lenn + lenp + 2"| rather than a
reference to \texttt{tmp}. The classical \texttt{NarrowArithAtSink}
body joins through \texttt{ActualArg.var $\to$ ResolvedVarType.var
$\to$ ArithOp.dst}, all three of which fail in this shape.
Pass~6 (\texttt{source\_arith\_sink\_bridge.dl}) closes the gap
with one rule that joins \texttt{ArithOp} and \texttt{Call}
directly when the call sits within two lines after the arith and
the callee is on a small alloc-class whitelist (xmlMalloc
family, av\_malloc family, libc allocators). In fewer than thirty
lines of Datalog, this recovers the libxml2 detection
(\S\ref{sec:eval-cve}).

\subsubsection{On-demand Datalog queries during reasoning}
\label{sec:m-ondemand}

The five-pass mesh runs once and produces a finding catalog
the host agent reads. That alone is a batch architecture:
the agent narrates verdicts from what the mesh materialised.
What it adds is the ability to ask new questions of the same
fact base \emph{during} reasoning --- ``do other functions share
this shape?'', ``does this intermediate carry a
\texttt{ResolvedVarType}?'', ``which sister sites in the same
module also fire?''. \sys exposes the loop through a single
on-demand Datalog tool.

\paragraph{The runtime.}
\texttt{neurolog\_\_query\_datalog} accepts arbitrary Datalog
source, a facts directory, the output relations to read back,
and optional pre-derived relations to stage (\texttt{ResolvedVarType},
\texttt{TaintedVar}, \texttt{BlockReach} from a prior batch run
as TSV CSVs). Souffl\'e\xspace runs in a temp directory with
the original facts symlinked in (no copy of 200K-row corpora)
and extras as \texttt{<rel>.facts}. The return is structured:
\texttt{ok} (rows, capped at 500/relation), \texttt{error}
(souffl\'e\xspace stderr plus a \emph{line-numbered} echo of
the rule text so \texttt{file:line:col} diagnostics align with
the agent's own source), \texttt{timeout}, or
\texttt{no\_outputs}.

\paragraph{What this changes.}
The host agent's role moves from ``read CSVs, write
verdict'' to ``read CSVs, hypothesise, query, refine, write
verdict.'' On the libxml2 case study (\S\ref{sec:eval-cve}) the
agent fired 19 ad-hoc queries in one investigation from a
one-paragraph user prompt. Two had schema mistakes that
Souffl\'e\xspace rejected, and the agent read the line-numbered
echo, fixed them, and resubmitted. The prior-art lineage of
``cheap static analysis narrows SMT'' (Saturn, Pinpoint,
Snugglebug) does not include this loop: those systems produce a
finding list and ask SMT for a yes/no per finding. The reasoning
agent with its own query tool over the finding list is what this
paper adds (eval: \S\ref{sec:eval-cve}).

\subsection{Stage 4 --- symbolic execution post-pass}
\label{sec:m-symbex}

For each Datalog finding $(\,f, \mathit{addr}, v, k\,)$, the
symbex stage encodes the def-use chain leading to $v$ at
$\mathit{addr}$ and the path guards along the way as a Z3
formula, asserts the bug condition for the bug class $k$, and
asks \texttt{check()}.

\paragraph{Phase A --- intra-procedural.}
The encoder walks back from the use of $v$ at $\mathit{addr}$ to
the most recent reaching def, encoding \texttt{ArithOp} on its
inputs recursively, \texttt{Cast} (zero/sign-extension or
truncation), and leaving \texttt{Call} return values free. Path
guards are collected by walking forward from the last common
ancestor in the CFG. For CVE-2023-45676: $v=\texttt{\_t}$ at 3653
is \texttt{ArithOp(+, len, 1)}; \texttt{len} at 3652 is
taint-source-defined and left as a free 32-bit \texttt{BitVec}.
The \texttt{narrow\_arith\_at\_sink} assertion is
\verb|len + 1 < len| as 32-bit unsigned --- satisfiable with
\texttt{len = 0xFFFFFFFF}.

\paragraph{Phase B --- function summaries + depth-bounded inlining.}
When dataflow crosses a function call, Phase~A's free symbol
misses information the callee provides --- a validator returns 0
or 1, an allocator returns non-null, a bitstream reader returns
at most $2^k-1$. Phase~B precomputes a small per-function summary
(return-value range, write-set, predicate over inputs) and
Phase~B-2 inlines callee bodies up to depth $d$ (default
$d{=}3$) when the summary is not specific enough.
Function-prefixed SSA keys prevent local-name collisions on
inline.

\paragraph{Phase D --- parallel verification.}
Findings never share encoder state, so the work is naturally
data-parallel: one worker per CPU, the \texttt{FactStore} loaded
once per worker, findings dispatched in chunks
($\approx$4 chunks/worker). On 378 ffmpeg-demuxer findings with
four workers, wall time drops from 1.19~s serial to 0.60~s
($1.98\times$); on 260 mov-demuxer findings, $2.55\times$. The
4--8-worker plateau is per-worker fixed cost (FactStore
deserialisation $\approx$50~ms, Python imports); pure Z3 scales
linearly past it.

\paragraph{Verdict and SAT model.}
\texttt{check()} returns \texttt{sat}, \texttt{unsat}, or
\texttt{unknown}. Feasible findings carry the model
(\verb|len_3652 = 4294967295, _t_3653 = 0, ...|) which both
accompanies the report and feeds the synthesis agent
(\S\ref{sec:m-synth}). \texttt{unsat} findings drop to a
secondary tier with a back-pointer to the excluding guard.

\subsection{Stage 4b --- likely range invariants (Phase E2)}
\label{sec:m-invariants}

Phase~A and B see only the source. They do not see what values
the program has actually held.
Sahoo~\textit{et al.}~\cite{sahoo2013likely} observe that 8--16
normal-input runs give tight range bounds on hot-function
locals, usable as a fault-localisation aid. We use the same
observation as a precision filter on SMT.

\paragraph{An optional dynamic extension.}
This is the first point where \sys \emph{runs} the target rather
than reading it, so it is worth being precise about what changes
and what does not. Stages~1--4 are the static core: no build, and
a complete tiered finding catalog out the other side. Stages~4b
and~5 are an \emph{optional} extension whose headline purpose is a
reproducible proof-of-vulnerability --- Stage~5 emits an actual
crashing input, which unavoidably runs the code --- and, having
paid for a runnable target, Stage~4b reuses the same machinery to
sharpen SMT ranges from observed values. Both go through a single
artifact built once per target: an AddressSanitizer-instrumented
executable taking one input file (often the project's existing
libFuzzer harness), at \texttt{-O0\,-g} (Figure~\ref{fig:pipeline},
stage~6). It is the only thing this extension compiles, and the
static core never does. Crucially, the extension can only
\emph{re-tier} findings, never erase them (tier-don't-drop,
\S\ref{sec:architecture}): with no harness the analyst still gets
the full static catalog, and with one, recall is unchanged and
only the ranking sharpens.

\paragraph{Mechanism.}
The runtime data we need is what \emph{normal} executions did at
the Datalog-flagged program points. We collect it by running that
harness on a corpus of valid inputs under GDB with a
breakpoint at every \texttt{(func, addr)} in the finding
catalog. A Python hook enumerates integer-typed locals via
\texttt{frame.block()} and prints one
\verb|INV|func|addr|var|val| line per local. After the corpus
pass we aggregate \texttt{(min, max, n\_obs)} per tuple into
\texttt{LikelyRangeInvariant(func, var, addr, lo, hi, n\_obs)}.
\texttt{encode\_var(v, def\_addr)} asserts
$\mathit{lo}\le v\le\mathit{hi}$ when a matching invariant exists.

\paragraph{Two-tier matching.}
Tier~1 is the exact triple \texttt{(func, def\_addr, var)};
Tier~2 widens to \texttt{(func, *, var)} (same variable anywhere
in the function, union of observed ranges). Tier~2 compensates
for compiler line-folding under \texttt{-O0+}. We deliberately
do \emph{not} accept a Tier~3 matching same-named variables
across functions: that would import a length field from one
parser as a bound on an unexercised attacker input in another,
silently demoting exactly the findings an analyst wants to keep.

\paragraph{Engineering notes.}
The harness must be \texttt{-O0 -g}, and LSan must be off
(\texttt{LSAN\_OPTIONS=detect\_leaks=0}, since it does not work
under ptrace). Multi-file projects read
\texttt{per\_function\_report.json}, so each breakpoint resolves
against its actual source file. On
the FFmpeg demuxer subset, three seeds (a 163-byte synthetic
Matroska scaffold and two truncations of a real Matroska file)
yield 245 invariants from 18\,602 observations across 348
breakpoints. Phase~B with invariants enabled demotes 13.2\,\%
of the FFmpeg feasible set, including the
\texttt{matroska\_parse\_webvtt} static FP that synthesis spent
five attempts on.

\subsection{Stage 4c --- dependence filtering and parser progress}
\label{sec:m-dep}

Two more lightweight precision passes attach to existing
boundaries.

\paragraph{E1 --- dependence filtering.}
The catalog often contains downstream echo findings (a tainted
index drives a finding at both the assignment and the sink).
Sahoo~\textit{et al.}'s dependence
clustering~\cite{sahoo2013likely} groups by data-flow
predecessor; we build a forward def-use graph keyed by
\texttt{(var, addr)} and BFS-reach each finding's source set.
Findings whose source set is properly contained in another's are
demoted to a \emph{downstream\_symptom} tier with a back-pointer
to the root. Tier annotation only --- nothing is removed.

\paragraph{E3 --- parser progress.}
On no-crash, synthesis feeds the previous attempt's ASan
transcript back into the prompt. To focus the next round on
\emph{how far} into the parser the attempt got, we count
distinct user-frame functions in the transcript (skip-list:
\texttt{main}, \texttt{LLVMFuzzerTestOneInput}, libc) and add
``previous attempt reached frames $A, B, C$; aim deeper'' to the
next prompt.

\subsection{Stage 5 --- crash-input synthesis (Phase C)}
\label{sec:m-synth}

\begin{figure}[t]
\centering
\scalebox{0.92}{%
\begin{tikzpicture}[
  node distance=3.5mm and 6mm,
  font=\scriptsize,
  inp/.style ={draw, rounded corners=2pt, fill=blue!4,
               inner sep=4pt, align=center, minimum width=22mm,
               minimum height=8.5mm, font=\scriptsize\ttfamily},
  llm/.style ={draw, rounded corners=2pt, fill=orange!12,
               inner sep=4pt, align=center, minimum width=24mm,
               minimum height=10mm},
  oracle/.style={draw, rounded corners=2pt, fill=blue!8,
               inner sep=4pt, align=center, minimum width=22mm,
               minimum height=10mm},
  good/.style={draw, rounded corners=2pt, fill=green!18,
               inner sep=4pt, align=center, minimum width=22mm,
               minimum height=8.5mm, font=\scriptsize\bfseries},
  arrow/.style={-{Stealth[length=1.6mm]}, thin}
]
  \node[inp]               (sat) {SAT model\\\textit{Z3 witness}};
  \node[inp, below=of sat] (sca) {scaffold ($P_c$)\\\textit{good prefix}};
  \node[inp, below=of sca] (vc)  {bug condition ($V_c$)\\\textit{e.g. len+1 $>$ 2\,GB}};

  \node[llm, right=14mm of sca] (agent)
       {Phase~C agent\\($N$ Python emitters\\per round)};

  \node[inp, right=12mm of agent, yshift=10mm,
        minimum width=22mm, font=\scriptsize\ttfamily]
        (cand) {N candidate blobs};

  \node[oracle, below=10mm of cand] (asan)
       {ASan harness\\\textit{ground truth}};

  \node[good, right=10mm of cand]    (crash) {crash\\(stop)};

  \draw[arrow] (sat.east) -- (agent.west|-sat.east);
  \draw[arrow] (sca.east) -- (agent.west|-sca.east);
  \draw[arrow] (vc.east)  -- (agent.west|-vc.east);

  \draw[arrow] (agent.east|-cand.west) -- (cand.west);
  \draw[arrow] (cand.south) -- (asan.north);
  \draw[arrow] (asan.east|-crash.west) -- (crash.west)
       node[midway, above, font=\tiny\itshape] {exit non-zero};

  \draw[arrow,bend right=20] (asan.south west) to
       node[midway, below, align=center, font=\tiny\itshape]
       {ASan tail +\\E3 progress score} (agent.south east);
\end{tikzpicture}}
\caption{Phase C multi-shot synthesis loop. The agent receives
the SMT witness (the value the bug condition needs), the
scaffold (a known-good $P_c$ prefix the parser will accept),
and the bug-class predicate $V_c$. It emits $N$ Python programs
per round whose outputs are validated against the AS\-an
harness. On no crash, the per-candidate AS\-an tail plus
the Phase~E3 parser-progress score feed back into the next
round so the agent can target deeper into the parser. The loop
terminates on the first crashing candidate or on a configurable
round budget (default 5).}
\label{fig:synth-loop}
\end{figure}
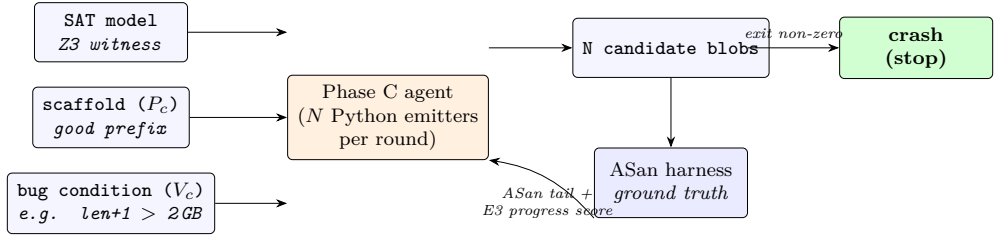

Synthesis takes a Phase~B-feasible finding plus its SAT model
and asks an LLM agent to write a Python program producing a
candidate crashing input. Multi-shot loop: $N$ candidates per
round (default $N{=}3$), each validated against the harness; on
no-crash, the per-candidate ASan tail and parser-progress score
feed the next round (Figure~\ref{fig:synth-loop}).

\paragraph{Prompt structure: $P_c$ + $V_c$.}
We decompose synthesis into a \emph{path condition} $P_c$ (a
working file-format prefix the parser accepts --- Vorbis comment
packet, MKV \texttt{D\_WEBVTT/SUBTITLES} track header) and a
\emph{vulnerability condition} $V_c$ (the constraint on the
trigger variable at the bug site --- e.g.\ the size argument
\texttt{(len+1)} exceeds the allocator limit, satisfied at
$\texttt{len}\approx\texttt{0x7FFFFFFF}$). $P_c$ is
handed to the agent as a \emph{scaffold} of bytes not to
disturb; $V_c$ as an explicit ``Target bug condition'' section.
The Python emitter is encouraged to produce a small scaffold
mutation, not a from-scratch generator. COTTONTAIL's
Constraint/Flexible-Mask decomposition~\cite{tu2026cottontail}
solves the same split at concolic-execution time; we solve it
statically.

\paragraph{Sanitizer harness as ground truth.}
Synthesis validates each candidate against the same sanitizer
harness introduced in \S\ref{sec:m-invariants} --- \sys's one
built artifact. A candidate is a crash iff the harness exits
non-zero with an ASan-recognisable signature.

\paragraph{Calibration on CVE-2023-45676.}
Phase~A proves the comment-length value reaches
\texttt{setup\_malloc} along a feasible path. The synthesis agent
recognises Vorbis from context --- it knows the Ogg
\texttt{capture\_pattern} and the comment-packet header. Its first
witness sets $\texttt{len} = \texttt{0xFFFFFFFF}$, but
\texttt{(len+1)} wraps to a zero-byte request that does not fault.
The no-crash ASan tail feeds back, and iteration~2 sets
$\texttt{len} = \texttt{0x7FFFFFFF}$, whose size argument (in
64-bit \texttt{size\_t}) is a $\sim$2\,GB request. The resulting
102-byte blob crashes ASan deterministically with
\texttt{allocation-size-too-big} at the sibling site
\texttt{start\_decoder:3653}. End-to-end synthesis is $\sim$5\,min,
almost all of it network round-trips, and no fuzzer is in the loop.

%% file: sections/06_implementation.tex
\section{Implementation}
\label{sec:implementation}

\sys is about 8\,KLoC of Python on top of the existing rules of
the sibling Datalog project, which we take unmodified. The module
layout follows the six pipeline stages of \S\ref{sec:architecture}.

\paragraph{Languages and dependencies.}
Tree-sitter, through the official \texttt{tree-sitter-c}
grammar~\cite{tree-sitter}, drives stage~1 and the mechanical
floor of stage~2. The grammar gives us language-aware identifier
extraction and CFG-edge enumeration without a working build of the
target. Souffl\'e\xspace 2.4~\cite{jordan2016souffle} runs the rule
mesh in stage~3. We invoke it as a subprocess and read its CSV
outputs directly. Z3 4.12~\cite{demoura2008z3} is used through the
standard Python bindings in stage~4. AddressSanitizer (Clang
\texttt{-fsanitize=address})~\cite{asan} is the oracle in stage~6,
wrapped by libFuzzer~\cite{libfuzzer} for single-input replay, or
by an equivalent thin C stub for projects that ship no fuzzer
harness.

\paragraph{LLM access and orchestration.}
Both in-pipeline LLM stages use the LiteLLM abstraction, so the
analysis is provider-neutral. In our experiments the smell pass
(stage~2) runs on a Lite-tier model (DeepSeek V4-Flash, through an
OpenAI-compatible endpoint) and the synthesis agent (stage~5) on a
frontier model (DeepSeek V4-Pro). Orchestration uses no bespoke
agent framework. \sys exposes its stages and on-demand queries as
Model-Context-Protocol (MCP) tools, and a \emph{host agent} --- a
coding agent such as Claude Code or OpenCode --- drives them,
interleaving \sys's deterministic tools with its own reasoning and
persisting verdicts back through the same interface. Any
MCP-capable host, and any model with an OpenAI-compatible API,
drops in.

\paragraph{Parallelism.}
The smell pass parallelises across functions with a thread pool
(15-way by default, since the bottleneck is provider throttling
rather than local CPU). Stage~4 parallelises across findings with
a process pool (\S\ref{sec:m-symbex}), and each worker loads the
\texttt{FactStore} once and processes a chunk of findings to
amortise startup. The synthesis agent runs $N{=}3$ candidate
emissions per round in parallel and validates each candidate
sequentially against the harness, since the harness is
single-threaded by default.

\paragraph{Caching.}
The smell pass writes a per-function cache keyed by a hash of the
source, the prompt template, and the model name. A re-run with the
same inputs reads from cache, which keeps iterative analyst work
cheap. The Datalog and SMT stages re-run on every change, since
they are seconds-scale and a full re-run is the simplest
dependency contract.

\paragraph{Per-scan audit log.}
Every component that performs a user-visible step (mech-extract
per file, smell-pass aggregate, Souffl\'e per output relation,
agent ad-hoc query, symbex per-batch, Phase~C per candidate) emits
one tab-separated line into a per-scan log when the
\texttt{NEUROLOG\_AUDIT\_LOG} environment variable points to a
path. The runners set it automatically per eval directory and
leave it off on interactive sessions to avoid noise. The log is
opt-in, threadsafe, append-only, and fail-safe: an audit-write
error never breaks the pipeline. Each line carries the TSV columns
\texttt{<ISO-timestamp>, <phase>, <action>, <target>, <key=value
detail>}, in a format meant to be parsed. This is the record of
what the tool actually did on a given run --- which Datalog
relations fired and with how many rows, which agent queries
returned what, which symbex batches landed where --- and it is
what makes a post-hoc audit possible without rerunning the
pipeline.

\paragraph{Artifact.}
The pipeline source, the rule mesh, the prompts, and the eval
scripts are released as an artifact. Reproducing the \texttt{stb}
CVE rediscovery and the FFmpeg precision tables of \S\ref{sec:eval}
needs only Souffl\'e, Python, and an LLM API key. Reproducing the
Phase~C crash synthesis additionally needs the ASan harness built
once: a one-line invocation on \texttt{stb}, and a documented
configure flag on FFmpeg.

%% file: sections/07_evaluation.tex
\section{Evaluation}
\label{sec:eval}

We evaluate \sys along five axes:
(i) end-to-end CVE rediscovery (\S\ref{sec:eval-cve}),
(ii) cost and recall of the LLM smell pass against an
LLM-only fact-extractor baseline (\S\ref{sec:eval-cost}),
(iii) precision of the symbex post-pass and the Phase E
filters (\S\ref{sec:eval-prec}),
(iv) wall-clock scaling of the parallel symbex stage
(\S\ref{sec:eval-d}), and
(v) end-to-end crash-input synthesis (\S\ref{sec:eval-c}).
We then summarise the honest negatives the evaluation surfaced
(\S\ref{sec:eval-neg}).

The evaluation targets are listed in
Table~\ref{tab:targets}.  All experiments are run on a single
laptop-class machine (5 CPU cores available to the pool) without
any accelerator. The two LLM stages call out to remote
provider APIs over the network, so the wall-clock cost on the LLM
side is dominated by provider round-trip rather than local
compute.

\begin{table*}[t]
\centering
\footnotesize
\begin{tabular}{@{}llp{55mm}rr@{}}
\toprule
Target dir & Codebase & Role & Lines & Fns sliced \\
\midrule
\texttt{stb\_legacy\_llm}     & stb\_image+truetype+vorbis & LLM-only-extractor cost baseline                & 18.6\,K & 128 \\
\texttt{stb\_mechanical\_smell} & same slice                & Cost-and-recall comparison (mechanical+smell)   & 18.6\,K & 128 \\
\texttt{stb\_fuzz\_validation} & stb\_vorbis                & Independent libFuzzer validation (23 unique crashes) & ---    & --- \\
\texttt{cjson}                 & cJSON v1.7.17              & 3-CVE rediscovery point                          & 2.5\,K  & 31  \\
\texttt{libxml2}               & libxml2~\cite{libxml2} (40 core .c files) & Scale demonstration + partial CVE rediscovery    & ---     & 2\,707 \\
\texttt{ffmpeg\_h264\_801}     & FFmpeg 8.0.1 H.264 slice   & Sentinel-collision case study (real CVE)         & ---     & 8 \\
\texttt{ffmpeg\_matroskadec}   & matroskadec.c              & Single-file demuxer eval                         & 5.1\,K  & 69 \\
\texttt{ffmpeg\_mov}           & mov.c                      & Sibling demuxer                                  & ---     & 153 \\
\texttt{ffmpeg\_demuxer\_full} & matroska + mov combined    & Combined demuxer eval (largest target)           & ---     & 222 \\
\bottomrule
\end{tabular}
\caption{Evaluation targets. The \texttt{stb} subdirectories
share the same source slice but exercise different extractor
configurations; the \texttt{ffmpeg} subdirectories cumulate to a
combined demuxer run.}
\label{tab:targets}
\end{table*}

\subsection{End-to-end CVE rediscovery}
\label{sec:eval-cve}

Table~\ref{tab:cves} lists every CVE \sys re-discovered
end-to-end in this evaluation.  ``End-to-end'' means: the
pipeline produced a Datalog finding pointing at the same source
line as the published CVE without that CVE being supplied as
input. Validation tier is one of three:
\textit{fuzzer-confirmed} --- an independent libFuzzer/AS\-an
run on the project produced a crash matching the static
finding's site; \textit{static-confirmed} --- the rule
mesh's evidence chain (taint source $\to$ guards $\to$ sink, with
relations cited) matches the published CVE writeup; or
\textit{partial} --- the pattern is surfaced but a CVE-class
semantics is not modelled (e.g.\ DoS, no memory-corruption
signal).

\begin{table*}[t]
\centering
\footnotesize
\renewcommand{\arraystretch}{1.15}
\begin{tabular}{@{}p{36mm}p{18mm}p{40mm}p{55mm}p{22mm}@{}}
\toprule
CVE & Codebase & Site & Detection rules (Datalog) & Validation \\
\midrule
CVE-2023-45681 (GHSL-2023-171)~\cite{cve-2023-45681}
   & stb\_vorbis
   & \texttt{setup\_malloc(sizeof(char*) * comment\_list\_length)}
   & \texttt{TaintedNarrowArith}, \texttt{TaintedSignedAtSink},
     \texttt{TaintedSizeAtSink}, \texttt{UnguardedTaintedSize}
   & Fuzzer-confirmed \\
CVE-2023-45676 (GHSL-2023-166)~\cite{cve-2023-45676}
   & stb\_vorbis
   & \texttt{setup\_malloc(sizeof(char) * (len+1))}
   & same four relations
   & Fuzzer-confirmed + Phase~C synth \\
CVE-2025-57052~\cite{cve-2025-57052}
   & cJSON 1.7.17
   & 4-fn taint chain $\to$ OOB array access
   & \texttt{UnguardedTaintedSink} (interproc)
   & Static-confirmed \\
CVE-2023-53154~\cite{cve-2023-53154}
   & cJSON 1.7.17
   & \texttt{parse\_string} heap over-read
   & \texttt{BufferOverflowInLoop}
   & Static-confirmed \\
CVE-2023-26819
   & cJSON 1.7.17
   & \texttt{parse\_number} DoS
   & \texttt{Cast} + tainted loop bound
   & Partial \\
FFmpeg H.264 sentinel collision
   & FFmpeg 8.0.1
   & \texttt{decode\_nal\_units} + \texttt{ff\_h264\_frame\_start}
   & \texttt{TaintedBuffer} + \texttt{ImplicitTruncation} +
     \texttt{UnboundedCounter} + \texttt{TaintControlledSink}
     (LLM-composed)
   & Static-confirmed \\
CVE-2023-38545~\cite{cve-2023-38545} (CVSS~9.8)
   & curl 8.3.0
   & \texttt{do\_SOCKS5} async-yield: hostname OOB write past
     \texttt{state.buffer}
   & \texttt{UnboundedServerWrite} (project-specific Pass);
     \texttt{JointBufferBoundUnsafe}
   & ASan-confirmed (T2) \\
curl SOCKS4 OOB~\cite{curl-socks-fix-8.4.0}
   & curl 8.3.0
   & \texttt{do\_SOCKS4} SOCKS4a: \texttt{strcpy} joint-bound miss
     (\texttt{plen}+\texttt{hostnamelen}>\texttt{buffer\_size})
   & \texttt{JointBufferBoundUnsafe} (project-specific Pass)
   & ASan-confirmed (T2) \\
CVE-2025-10148~\cite{cve-2025-10148}
   & curl 8.3.0
   & \texttt{ws\_enc\_write\_payload} XORs every frame through a
     once-generated mask
   & \texttt{SingleWriterSecurityField} (Pass~11)
   & Static-confirmed \\
curl WS heap OOB read
   & curl 8.3.0
   & \texttt{ws\_dec\_read\_head} accepts MSB-set 64-bit
     \texttt{payload\_len} $\to$ signed-to-unsigned OOB in
     \texttt{ws\_dec\_pass\_payload}
   & \texttt{TaintedSignExtension} +
     \texttt{TaintedWidthMismatchAtSink}
   & ASan-confirmed (T2) \\
curl WS busy-loop DoS
   & curl 8.3.0
   & \texttt{ws\_flush} non-destructive peek + EAGAIN-continue
     without buffer advance
   & \texttt{NondestructivePeekInLoop} (Pass~14)
   & Static-confirmed \\
\bottomrule
\end{tabular}
\caption{CVE rediscovery ledger.  \sys re-discovered eight fully
+ one partial CVE-class issues end-to-end across the codebases
evaluated, including a CVSS-9.8 \texttt{curl} SOCKS5 vulnerability.
The two \texttt{stb} CVEs are additionally confirmed by independent
libFuzzer/AS\-an runs producing 23 unique crashes that cluster at
the static-finding sites; the two \texttt{curl} OOB rows are
confirmed by 5-line ASan harnesses that fire deterministically
against the un-patched 8.3.0 source.}
\label{tab:cves}
\end{table*}

The two \texttt{stb} CVEs are independently confirmed by a
libFuzzer/AS\-an run on \texttt{fuzz\_vorbis} producing 23
unique crashes clustered in the \texttt{stb\_vorbis} decode
chain at the lines Datalog flagged
($L_{3664/3670}\to L_{5112}\to L_{5390}$). The FFmpeg H.264
sentinel-collision is notable because the compound pattern is
deliberately not encoded as one rule; the mesh produces four
ingredient relations and the host agent composes them.
The libxml2 run is a scale demonstration: 40 source files,
2\,707 functions, 185\,839 facts, full pipeline 73\,min
(smell pass 22.6\,min, Souffl\'e\xspace 16.7\,min, symbex
Phase~A returns 2\,226 feasibles). The smell pass agrees with
the mechanical extractor on almost all facts (1 addition, 0
corrections in 2\,707 calls).

We also tested whether the mesh re-discovers
CVE-2025-6021~\cite{cve-2025-6021} in libxml2's
\texttt{xmlBuildQName} and four sister sites, this time using
the pipeline as intended: handing the artefacts to the
host agent with the on-demand
\texttt{neurolog\_\_query\_datalog} tool (\S\ref{sec:m-ondemand})
and a one-paragraph prompt. Across 89 events the agent fired
\textbf{19 ad-hoc Datalog queries} that authored their own
schemas, joined precomputed relations against raw facts, and
recovered from two schema-arity errors after reading
Souffl\'e's line-numbered echoes. The agent's diagnosis was
sharper than ours: the affected set is six functions, not four
(two more recovered by enumerating \texttt{ArithOp} facts with
destination \texttt{tmp} and operands $\{$\texttt{lenn},
\texttt{lenp}$\}$); the recall miss is compound, not
single-rule (LLM-introduced \texttt{tmp} carries no
\texttt{ResolvedVarType}; \texttt{ActualArg} records the
literal expression string, not a reference to \texttt{tmp}).
The fix is a small bridging rule plus an extraction-layer
patch, shipping as Pass~6
(\texttt{source\_arith\_sink\_bridge.dl}, \S\ref{sec:m-datalog}).
Re-run on the same fact base, Pass~6 recovers
\textbf{6 of 6} CVE-2025-6021 sites in 181 candidate (function,
sink) pairs across 2\,707 functions; applied to FFmpeg facts
the same rule surfaces 41 candidates on the combined demuxer
corpus and generalises to unrelated targets like
\texttt{matroska\_decode\_buffer:1707}. The on-demand-query
architecture is earning its keep.

\paragraph{CVE-2023-38545: rule developed end-to-end.}
The two \texttt{curl} SOCKS rows share a structural pattern no
existing rule covered: separate bounds checks on two operands
(\texttt{plen}, \texttt{hostnamelen}) each individually sound,
but whose \emph{joint} bound vs.\ destination capacity is never
asserted. We encoded the pattern as
\texttt{joint\_buffer\_bound.dl}; on curl 8.3.0 it fires at
\texttt{do\_SOCKS4:430} (the SOCKS4a strcpy upstream patch
\texttt{01057d6161} later guarded) and at the
\texttt{do\_SOCKS5:907} sister site that became
CVE-2023-38545. Both fires reach the synthesis stage and ASan
confirms a 3-byte overflow for the SOCKS4a path and a 2000-byte
overflow for the CVE-2023-38545 path (the latter requires
driving the async-yield state-machine bug that resets the local
\texttt{socks5\_resolve\_local} flag on re-entry; an 800\,ms
stub delay suffices). This is the first eval target where we
wrote a new rule \emph{because of} the audit and watched it
generalise: the do\_SOCKS5 fire was not the bug we set out to
find.

\paragraph{WebSocket: mesh localises, LLM reads the neighbourhood.}
Auditing \texttt{lib/ws.c} re-discovered 6 of the 13 fixes that
landed upstream between 8.3.0 and HEAD; beyond the three in
Table~\ref{tab:cves} the system also surfaced a wrong-offset
multi-fragment \texttt{memcpy} (silently fixed in \#12945),
missing RSV-bit validation (\#16069), and an uninitialised
send-count leak in \texttt{Curl\_senddata}. Tellingly, the last
two came not from a rule fire but from the LLM reading the
decoder a mesh fire had already pointed it at --- the
neuro-symbolic loop working as intended. The 6 misses cluster
into two classes the mesh did not yet cover, cross-function
state-machine invariants and resource-cleanup paths, which
directly motivated Pass~15 (\texttt{state\_machine\_invariants.dl})
and Pass~16 (\texttt{lifecycle\_audit.dl}).

\paragraph{Recon front-end: a credential-scope case study.}
The two miss classes above are exactly what the recon front-end
(\S\ref{sec:m-recon}) targets. As a capability check we pointed recon
at \texttt{curl}'s connection-reuse logic (\texttt{lib/url.c}), which
the sink-seeded slice never reaches. Recon proposed the reuse
comparators as seeds and a \emph{connection-reuse credential-scope}
hypothesis with a Datalog check; the smell pass then tagged each
compared field via \texttt{FieldSemantic}, correctly recognising
\texttt{Curl\_timestrcmp(needle->user, check->user)} as a credential
comparison. The omission rule fired on exactly one site ---
\texttt{proxy\_info\_matches}, which gates proxy-connection reuse on
host/port/type but never on the proxy \texttt{user}/\texttt{passwd} the
struct carries --- and correctly cleared \texttt{ConnectionExists}
(which does compare its credentials) and an age-check helper. This is
the connection-reuse credential class that dominates the 2026
\texttt{curl} advisory wave, a class no memory-safety rule models. We
report it as a demonstration of the perceive-then-prove loop on a
non-memory class, not as a novel finding (the underlying issue is
plausibly a rediscovery), and we leave a broad multi-target evaluation
of the front-end to future work.

\subsection{Novel-bug discovery on libarchive HEAD}
\label{sec:eval-libarchive}

To test whether the mesh surfaces \emph{novel} issues on
current code, we audited libarchive master at commit
\texttt{a651b4fc} (HEAD, 2026-05-19) across five formats
(XAR/WARC/7zip writers; cpio and iso9660 readers). The mesh's
\texttt{LifecycleAudit}, \texttt{ResourceCleanupMiss}, and a
host-authored \texttt{free-while-linked} lens produced five
candidate sites, all hand-validated under ASan/LSan: one
use-after-free + double-free in the cpio reader
(issue~\#3053~\cite{libarchive-issue-3053}); three resource-leak
bugs in the XAR reader, WARC writer, and 7zip
writer~\cite{libarchive-issue-xar-reader,
libarchive-issue-warc-writer, libarchive-issue-7zip-writer};
and one OOB read into adjacent \texttt{.rodata} in the XAR
writer's \texttt{make\_fflags\_entry} (issue~\#3059
\cite{libarchive-issue-xar-writer}).\footnote{Concurrent
discovery note. An unrelated reporter had independently filed
the same XAR-writer issue under private security advisory
\texttt{GHSA-wfvr-54j8-47r9}; the maintainer's fix
PR~\#3041~\cite{libarchive-pr-3041} was opened 2026-05-16,
roughly two days before our independent finding, and has since
been merged (issue~\#3059 closed). We learned about the
in-flight fix only after we filed issue \#3059, when the
maintainer pointed us at PR \#3041. Our reproducer fires
deterministically against HEAD as of our audit and is silenced
by PR~\#3041 (we verified). The maintainer's fix reorders the
inner comparison to gate the OOB read behind a successful
\texttt{strncmp}; our proposed fix replaces the null-terminator
trick with an explicit \texttt{strlen}-based length check.
Both are functionally equivalent for valid inputs.} All five
sites are filed upstream; per-bug source locations, ASan
witnesses, reproducers, and proposed patches are in the linked
issues.

The cpio finding is the most interesting individual result: a
regression that the upstream commit which introduced it (a leak
fix, 23 days earlier) made worse, not better. On opening the
issue, the maintainer acknowledged it and opened PR~\#3055,
crediting the reporter as commit author, with the fix following
our proposed shape. The XAR-reader leak finding
(issue~\#3058) was similarly acknowledged with fix PR~\#3060, whose body restates
our issue text verbatim. At submission, the 7zip-writer
(PR~\#3062) and XAR-reader (PR~\#3060) fixes are merged and the
XAR-writer OOB read is resolved upstream via PR~\#3041; the
WARC-writer (PR~\#3061, approved) and cpio (PR~\#3055) fixes
remain in review.

The audit yields a small taxonomy of ``where the mesh found
bugs fuzzing did not'': no test
exercises writer-side error paths (two leaks); cleanup that misses one of many struct
members because the populator lives in a sibling translation
unit (one leak); string-literal lookup tables whose fast
pre-filter trusts the null terminator (one OOB read);
ENOMEM-gated paths fuzzers under-explore (the cpio UAF). None
are reader-side parser bugs OSS-Fuzz would have surfaced
--- evidence that \sys complements fuzzing rather than competing
with it.

\subsection{Cost and recall of the LLM smell pass}
\label{sec:eval-cost}

The mechanical floor + smell pass is one of the two extractor
configurations in \S\ref{sec:m-facts}; the other is a
frontier-model-only extractor reading the full function source
without the tree-sitter floor. Both run on the same 128-function
\texttt{stb} slice (Table~\ref{tab:cost}).

\begin{table}[t]
\centering
\footnotesize
\begin{tabular}{@{}lrrr@{}}
\toprule
Metric & LLM-only & Mech.\,+\,smell & Ratio \\
\midrule
Wall time (extraction)        & 55 min           & 36.6 s   & $\approx 90\times$ \\
Cost (DeepSeek API)           & \$0.30--\$0.50   & \$0.005  & $\approx 70\times$ \\
Functions covered             & 106 / 128         & 128 / 128 & 1.21$\times$ \\
Critical-bug fns covered      & 0 / 3 (truncated) & 3 / 3    & --- \\
CVE-class findings            & 2                 & 2        & parity \\
\bottomrule
\end{tabular}
\caption{Cost and recall on the same \texttt{stb} 128-function
slice. The frontier extractor truncates the 627-line
\texttt{start\_decoder} function and so misses the bug-relevant
content; the mechanical+smell pipeline does not, recovering
recall on exactly the function CVE-2023-45676 lives in.}
\label{tab:cost}
\end{table}

$90{\times}$ faster, $70{\times}$ cheaper, recovers coverage on
three bug-relevant functions the frontier extractor truncated,
matches CVE-class recall. This is the empirical license for
\S\ref{sec:m-facts}: the LLM is best at function-local
augmentation, not whole-codebase reading.

\subsection{Symbex precision and Phase E filters}
\label{sec:eval-prec}

We measure precision at three points: Phase~A $\to$ B flips,
Phase E1 root-vs-symptom split, and Phase E2 invariant reduction.

\paragraph{Phase A $\to$ Phase B (Tab.~\ref{tab:phaseab}).}
Phase~A path-guard filtering does most of the work --- 28--38
infeasibles per FFmpeg target with no catalog. Phase~B summaries
flip exactly one finding overall: \texttt{slice\_type} in
\texttt{ffmpeg\_h264\_801}, where
\texttt{get\_ue\_golomb\_31}'s catalog summary
(return\,$\in[0,31]$) tightened the constraint. Demuxer callees
flow through \texttt{ebml\_*} / \texttt{avio\_*} variants we have
not catalogued, so summaries do not bind there.

\begin{table}[t]
\centering
\footnotesize
\begin{tabular}{@{}lrrrrr@{}}
\toprule
Target & DL findings & SMT-checked & Phase A f/i & Phase B f/i & B flips \\
\midrule
ffmpeg\_h264\_801      & 1\,064 & 180 & 145/35 & 144/36 & \textbf{1} \\
ffmpeg\_matroskadec    &   482 & 118 & 108/10 & 108/10 & 0 \\
ffmpeg\_mov            & 1\,361 & 260 & 232/28 & 232/28 & 0 \\
ffmpeg\_demuxer\_full  & 1\,843 & 378 & 340/38 & 340/38 & 0 \\
stb\_mechanical\_smell &    --- & 145 & 134/11 & 134/11 & 0 \\
\bottomrule
\end{tabular}
\caption{Phase A and Phase B verdict counts. ``f/i'' is
feasible/infeasible. ``B flips'' is the number of findings whose
verdict changed when summaries plus depth-bounded inlining were
added.  Phase~A is doing the bulk of the work; Phase~B's
contribution is targeted (where the codebase has spec-validated
bitstream readers).}
\label{tab:phaseab}
\end{table}

\paragraph{Phase E1 dependence filtering (Tab.~\ref{tab:e1}).}
Symptom share is lower than Sahoo et al.'s
58\,\%~\cite{sahoo2013likely}: their dataset was real-bug-centric
(single root, many symptoms); ours is broad static coverage with
many independent reads. The cluster structure remains useful:
\texttt{mov\_read\_header} has 7 symptoms behind 2 roots (2 bugs
to investigate, not 9); \texttt{matroska\_parse\_laces} has 1
symptom behind 6 roots (a tightly-coupled region readable
top-down).

\begin{table}[t]
\centering
\footnotesize
\begin{tabular}{@{}lrrrr@{}}
\toprule
Target & B feasible & Root cause & Symptom & Symptom \% \\
\midrule
ffmpeg\_h264\_801      & 141 & 129 & 12 & 8.5\,\% \\
ffmpeg\_matroskadec    &  90 &  86 &  4 & 4.4\,\% \\
ffmpeg\_mov            & 222 & 192 & 30 & 13.5\,\% \\
ffmpeg\_demuxer\_full  & 312 & 278 & 34 & \textbf{10.9\,\%} \\
stb\_mechanical\_smell & 134 & 131 &  3 & 2.2\,\% \\
\bottomrule
\end{tabular}
\caption{Phase E1 dependence-filter tier breakdown. ``Symptom''
findings are demoted with a back-pointer to the root; nothing is
removed.}
\label{tab:e1}
\end{table}

\paragraph{Phase E2 likely-invariants (Tab.~\ref{tab:e2}).}
On \texttt{stb}, eight Vorbis seeds give 43.3\,\% reduction,
all in \texttt{start\_decoder}'s loop counters; CVE-2023-45676
/45681 sites are unaffected because they live in relations
Phase~E2 does not consume (\texttt{SignConfusionCast},
\texttt{TypeSafetyFinding}). On FFmpeg demuxer, three Matroska
seeds give 13.2\,\%; relations asking about ``out-of-range
arithmetic'' are demoted the most (\texttt{NarrowArithAtSink}
80\,\%) because the corpus provides direct counter-examples.

\begin{table}[ht!]
\centering
\footnotesize
\begin{tabular}{@{}lrrr@{}}
\toprule
Target & Phase B feasible & B + E2 feasible & Reduction \\
\midrule
stb\_mechanical\_smell & 134 &  76 & \textbf{$-43.3\,\%$} \\
ffmpeg\_demuxer\_full  & 340 & 295 & \textbf{$-13.2\,\%$} \\
\bottomrule
\end{tabular}

\medskip

\begin{tabular}{@{}lrrr@{}}
\toprule
Relation (ffmpeg) & B-feasible & B+E2 feasible & Demoted \\
\midrule
\texttt{NarrowArithAtSink}      &   5 &   1 & \textbf{4 (80\,\%)} \\
\texttt{SignedArgAtSink}        &  12 &   9 & 3 (25\,\%) \\
\texttt{ImplicitTruncation}     & 140 & 123 & 17 (12.1\,\%) \\
\texttt{PotentialArithOverflow} & 183 & 162 & 21 (11.5\,\%) \\
\bottomrule
\end{tabular}
\caption{Phase E2 likely-invariant reduction. The matroska
\texttt{parse\_webvtt:3853} static FP (\S\ref{sec:eval-c}) is
one of the four \texttt{NarrowArith} demotions: the corpus
observed \texttt{text\_len}$\in[0,28]$ over 28 hits at L3842,
so the bug condition becomes UNSAT under the Tier-2 match.}
\label{tab:e2}
\end{table}

\subsection{Phase D --- parallel symbex scaling}
\label{sec:eval-d}

Table~\ref{tab:phased} reports wall-clock serial vs.\ at
$N{=}2,4,8$ workers (parity against serial verified).

\begin{table}[ht!]
\centering
\footnotesize
\begin{tabular}{@{}lrrrrr@{}}
\toprule
Target & $n$ findings & serial & par(2) & par(4) & par(8) \\
\midrule
ffmpeg\_h264\_801      & 180 & 0.38 s & 0.26 s & \textbf{0.19 s} (2.0$\times$) & 0.21 s \\
ffmpeg\_matroskadec    & 118 & 0.40 s & 0.28 s & \textbf{0.18 s} (2.2$\times$) & 0.20 s \\
ffmpeg\_mov            & 260 & 0.84 s & 0.57 s & \textbf{0.33 s} (2.6$\times$) & 0.36 s \\
ffmpeg\_demuxer\_full  & 378 & 1.19 s & 0.79 s & 0.60 s (2.0$\times$) & \textbf{0.52 s} (2.3$\times$) \\
\bottomrule
\end{tabular}
\caption{Phase D parallel symbex wall-clock. Best speedup per
target is bold. Plateau at 4--8 workers is per-worker
fixed-cost (FactStore deserialisation, Python imports); pure-Z3
work scales linearly past that plateau.}
\label{tab:phased}
\end{table}

The 4-worker plateau is per-worker fixed cost ($\approx 50$\,ms
FactStore + Python imports), not Z3 saturation: on
\texttt{ffmpeg\_demuxer\_full} the 8-worker speedup
($2.3\times$) does extend past 4-worker, leaving room at
codebase scale.

\subsection{Phase C --- crash-input synthesis}
\label{sec:eval-c}

\paragraph{Calibration on \texttt{stb\_vorbis}.}
On CVE-2023-45676, the Phase~B finding
\texttt{start\_decoder:3670 tmp unbounded\_counter\_at\_sink}
feeds the synthesis agent with the SMT model
($\texttt{vendor\_length}=$\texttt{0x7FFFFFFF}), the def-use
chain back to \texttt{get32\_packet}, the snippet, and a Vorbis
format hint. At $N{=}3$ candidates/round, iteration~2 produces
a 102-byte Ogg/Vorbis blob crashing \texttt{fuzz\_vorbis}
(ASan: \texttt{allocation-size-too-big} at
\texttt{start\_decoder:3653}, the sibling site of the same CVE
class). End-to-end $\approx$5\,min, mostly network round-trips.
Reproducible via
\verb|fuzz_vorbis stb_mechanical_smell/synth_crash.bin|. \emph{No
fuzzer was used in the synthesis loop}; the only fuzzer contact
is the ASan oracle in stage~6.

\paragraph{FFmpeg \texttt{matroska\_parse\_webvtt:3853} ---
honest negative.}
On the FFmpeg matroska demuxer's \texttt{text\_len} finding
(Phase~B feasible without invariants), five iterations of
progressively-better setups (matroska-specific harness,
scaffold seeding, $P_c/V_c$ prompts) produced 161--169-byte
mutations of the 163-byte scaffold; GDB confirmed
candidate~0 of synth~5 reaches the target function.
\emph{No crash.} Investigation showed a static FP:
\texttt{text\_len = q - p} where $q=\texttt{data}+\texttt{data\_len}$
and $p=\texttt{data}$, so $q < p$ is unreachable. This is the
case Phase~E2 was designed for: running E2 on a three-seed
Matroska corpus demotes the finding (Tab.~\ref{tab:e2}). The
closed-loop story: synthesis exhausted five attempts on a static
FP, the analyst recognised it, ran E2 once, the next pipeline
run no longer flags the site.

\paragraph{What this validates and what it does not.}
The framework is end-to-end functional on real-world FFmpeg ---
ASan-instrumented per-demuxer harnesses build cleanly, scaffold
synthesis produces correctly-structured mutations that reach
the target function, and the chain-walker fix following plain
assignment edges is permanent. What FFmpeg does \emph{not}
demonstrate is novel-CVE discovery on its own corpus: the
matroska finding was a static FP, and the other demuxer
feasibles remain to be triaged manually.

\subsection{Honest negatives}
\label{sec:eval-neg}

We summarise the negatives the evaluation surfaced so a reader
can calibrate the headline numbers. These negatives also constitute our future work in this direction.

\begin{itemize}[leftmargin=*]

\item \textbf{Fact-extraction recall on LLM-introduced
intermediates.} On the libxml2 \texttt{xmlBuildQName} site and
its five sister patterns in \texttt{xmlregexp.c}, the smell
pass emits an \texttt{ArithOp} for \texttt{tmp = lenn + lenp +
2} but no \texttt{Def} / \texttt{VarType} /
\texttt{ResolvedVarType} on \texttt{tmp}, and the call's
\texttt{ActualArg} records the literal expression string rather
than a reference to \texttt{tmp}. The headline \texttt{int +
int} arithmetic-overflow rule misses all six sites until the
bridging Pass~6 (\S\ref{sec:m-datalog}) is added. The
underlying gap --- the LLM does not type its synthetic
intermediates --- is the worst LLM-recall failure mode the
evaluation surfaced and the most plausible source of further
silent misses in the long tail.

\item \textbf{Rule-mesh coverage gaps.} The mesh does not
model: DoS-by-recursion, UAFs whose free is conditional on a
callee return, and compound patterns like FFmpeg's H.264
32-bit-counter-into-16-bit-sentinel collision (the four
ingredient relations exist; rule-level composition does not).
The H.264 case is handled by the host agent composing
relations after the fact; mesh recall on that family is bounded
by the agent's compositional ability (but this is also the design philosophy of~\sys --- on-demand datalog queries).

\item \textbf{Phase~B catalog effects are small.} The targeted
catalog of bitstream readers (\texttt{get\_ue\_golomb\_31} and
friends) flips one finding total in our evaluation. The
mechanism works --- the flip is genuinely the right call --- but
the catalog is small relative to the function space, so its
overall precision contribution is modest.

\item \textbf{Phase E1 symptom share is low.}
At 2\,--\,13\,\% across targets, much lower than
Sahoo~\textit{et al.}'s 58\,\%~\cite{sahoo2013likely}. The
likely cause is workload structure: their dataset was
real-bug-centric (single root, many symptoms); ours is broad
static coverage (many independent reads). The cluster
\emph{structure} is still useful for triage even when the
percentage is small.

\item \textbf{Phase~E2 ordering cost five synthesis iterations.}
We spent five Phase~C rounds confirming that
\texttt{matroska\_parse\_webvtt:3853} cannot crash before
recognising it as a static FP and running E2. Phase~E2 currently
runs \emph{after} Phase~C; running it first would have caught the
false positive before synthesis ever started. Re-ordering is
mechanical (\S\ref{sec:discussion}).

\item \textbf{Souffl\'e\xspace is the dominant non-LLM cost
at scale.} On libxml2 (185\,839 facts, 2\,707 functions), the
five-pass mesh takes 16.7\,min --- larger than the LLM smell
pass at 22.6\,min at 15-way parallelism, but the same order.
At codebase scales above \texttt{libxml2} the Souffl\'e\xspace
runtime will dominate; we have not investigated the standard
mitigations (table partitioning, incremental evaluation).

\item \textbf{No head-to-head against industrial baselines.}
We did not run CodeQL or Joern on the same targets. CodeQL
requires a working build (which our targets either lack or
require significant scaffolding for), and Joern's open-source
C/C++ frontend does not carry a runtime-grade decision
procedure for path feasibility. The precision and recall
claims in this paper are therefore absolute rather than
relative; an apples-to-apples comparison on Magma is the
natural follow-up benchmark (\S\ref{sec:discussion}).

\item \textbf{Triage backlog.} The smell pass produces several
hundred \texttt{possible\_uninit\_free} flags on FFmpeg that we
have not reviewed; the E2-filtered FFmpeg demuxer feasible set
(295 findings) is also untriaged. These remain the most
plausible source of additional high-confidence findings on the
codebases evaluated, but per-flag manual review is the
bottleneck we have not yet paid.

\end{itemize}

%% file: sections/08_discussion.tex
\section{Discussion}
\label{sec:discussion}

\subsection{Threats to validity}
\label{sec:disc-threats}

\paragraph{Silent recall misses from the LLM.}
The biggest threat is that the smell pass omits a fact the rule
mesh needs: a missing \texttt{Cast} silently kills an
\texttt{ImplicitTruncation} finding. We mitigate it three ways.
The mechanical floor produces structural facts deterministically,
so the LLM cannot omit them. The schema is small and
security-aware, which keeps the prompt focused. And
\S\ref{sec:eval-cost} matches CVE-class recall against a
frontier-model-only baseline on a held-out slice. None of these
prove that the LLM never misses in the long tail. The libxml2
\texttt{xmlBuildQName} miss is exactly this category.

\paragraph{Compile-free trade-offs.}
\sys gives up the precision a full build would buy. The
preprocessor is not run, so macros that conditionally insert
guards or sinks are invisible, and the type system is not lifted,
so struct-definition-driven disambiguation is unavailable. We pay
this knowingly, to gain the property that an analyst can run on an
unfamiliar codebase in seconds. A build-based front end could
extract the same facts more precisely. We claim only that the cost
difference is worth the precision gap on the codebases we
evaluated.

\paragraph{Rule-mesh coverage.}
The mesh is the union of two prior efforts: the sibling project's
binary-side rules plus source-side extensions. It does not cover
every CVE class. DoS-by-recursion, certain UAFs where the free is
conditional on a callee return, and the H.264 sentinel-collision
compound pattern are all under-covered. The H.264 case is honest:
four ingredient relations surface and the agent layer composes
them, so mesh recall there is bounded by the agent's compositional
ability on findings the mesh did not directly flag. The mesh is
also expandable by design, so an analyst can add rules from their
own target's audit.

\paragraph{Single LLM provider.}
Both LLM stages were exercised against DeepSeek V4 in our
experiments. The smell pass is generic to any OpenAI-compatible
endpoint, and the synthesis agent uses provider-neutral LiteLLM.
We have not run a head-to-head with other providers, so we do not
claim provider-independence. We claim only that the design uses no
provider-specific features: no fine-tuning, no internal APIs, no
proprietary tools.

\paragraph{Scale of the evaluation.}
Eight rediscovered CVE-class issues plus five novel libarchive
findings across six codebases is a moderate envelope.
Magma~\cite{hazimeh2020magma} is the natural next benchmark, but a
fuzzer-targeted benchmark --- manually-injected bugs,
crashes-found labels --- is structurally different from a
static-analyser evaluation, and we kept this paper's targets
comparable to the manual review practitioners do today.

\subsection{Design decisions revisited}
\label{sec:disc-design}

\paragraph{LLM as fact extractor, not query composer.}
The most natural alternative design uses the LLM at \emph{query}
time: the analyst describes a vulnerability pattern in natural
language, the LLM emits a Datalog query, the engine runs it, and
the analyst iterates. Such a design suits a setting where the fact
base already exists --- typically because a production analyser
has lifted the project to an IR --- and the analyst's bottleneck
is expressing the pattern, not extracting the facts. \sys takes
the opposite tack, because on unbuilt source the fact extraction
\emph{is} the bottleneck, and because a fact base is auditable in
a way an ad-hoc query is not: we can diff two runs and see which
tuples appeared or disappeared. We move the LLM \emph{below} the
rule mesh. It fills tuples, and the rules express patterns once
and reuse them across targets.

\paragraph{SMT as a post-pass, not in the rule body.}
We discuss this in \S\ref{sec:related} and \S\ref{sec:m-symbex}.
The deliberate trade is full Souffl\'e\xspace throughput,
per-finding parallelism, and an exposed SAT model that downstream
stages consume. Formulog's intra-rule integration buys earlier
pruning, at the cost of a bespoke evaluation strategy and an
unexposed model. The two designs are not strictly comparable on
benchmarks, since the workloads differ, but post-pass is the
simpler retrofit onto an existing rule corpus --- which is what we
needed.

\paragraph{Tier-don't-drop vs.\ aggressive filtering.}
Every Phase~E pass exposes a tier rather than removing findings,
which guards against mute-then-miss: an attacker input the corpus
did not exercise is exactly what an aggressive filter would mute.
We trade a longer finding list for correctness, assuming a report
UI that shows tier-1 by default and expands lower tiers on demand.

\subsection{Future work}
\label{sec:disc-future}

\paragraph{Re-order Phase E2 to run before Phase C.}
The matroska FP of \S\ref{sec:eval-c} cost five synthesis
iterations because Phase~E2 runs after Phase~C. Running E2 first
would have demoted the finding before synthesis ever started. The
re-ordering is mechanical and is the clearest single improvement
the evaluation pointed to.

\paragraph{Magma benchmark with patch/unpatch.}
Magma~\cite{hazimeh2020magma} ships curated real vulnerabilities
in libtiff, libxml2, libpng, poppler, and others, each toggled by
a one-line patch. The patched tree should give zero
high-confidence findings at the site (an FP rate), and the
unpatched tree should surface the finding (TP recall). A direct
comparison with classical static analysers and fuzzers on the same
labelled set is the obvious next experiment, and because Magma
supplies the harness, the only per-target variable is the sink
catalog.

\paragraph{Cross-codebase invariant transfer.}
Phase~E2 invariants are project-local. Whether range invariants
extracted on one codebase --- say, ``a length field after
\texttt{varint} decode is $\le 2^{32}-1$'' --- transfer to other
codebases in the same protocol family is open. If they do, the
seed cost amortises across projects.

\paragraph{Concolic-style $P_c$-byte mapping.}
Synthesis currently hands the SMT model to the LLM as text, and
the LLM maps model values back to byte offsets. COTTONTAIL's
runtime concolic tracing~\cite{tu2026cottontail} gives a more
direct mapping, at the cost of running the target. A static
approximation --- tracking each $P_c$ symbol back through the
parser to a byte offset, even imperfectly --- is within reach.

\paragraph{Richer rule-mesh coverage.}
Use-after-free with conditional frees, type-confusion across union
members, and DoS-by-recursion are all recall gaps we identified
during evaluation. Adding rules is mechanical. The limit is
analyst time, not theoretical complexity.

%% file: sections/09_related.tex
\section{Related work}
\label{sec:related}

\sys is most usefully positioned as a re-application of an old idea
in a new substrate. We organise this section around the four
threads of work it touches: (i) Datalog-based static analysis,
(ii) the Datalog\,+\,SMT combination specifically, (iii) the
broader ``cheap static analysis narrows SMT'' lineage, and (iv)
LLMs for vulnerability detection and structured input generation.

\paragraph{Datalog-based static analysis.}
The Datalog school of static analysis has matured into the
production tooling we treat as a substrate. Doop's points-to
analysis~\cite{bravenboer2009doop} demonstrated that sophisticated
context-sensitive pointer reasoning fits in a few hundred lines of
Horn clauses. Souffl\'e\xspace~\cite{jordan2016souffle} synthesises
parallel C++ from such rules and is the engine \sys uses
unmodified. CodeQL~\cite{avgustinov2016ql} packages a
Datalog-style query language for industrial security review, and
code property graphs~\cite{yamaguchi2014cpg} share Datalog's
auditability while emphasising graph-reachability patterns. \sys
reuses the \emph{rule mesh} idiom of these systems but moves the IR
producer --- the part that would normally consume LLVM, Java
bytecode, or a build-wrapper-instrumented C compiler --- to a
language model that reads source directly.

\paragraph{Datalog with SMT.}
Formulog~\cite{bembenek2020formulog} introduced an SMT-aware
Datalog dialect in which \texttt{is\_sat()} is callable from a rule
body, and the engine schedules SMT queries during semi-naive
evaluation. The 2024 follow-up~\cite{bembenek2024fast} shows that
an unconventional ``eager evaluation'' strategy gives $5{-}7\times$
speedups on SMT-heavy workloads. Formulog's three case studies
cover refinement-type checking, Java pointer analysis, and a
Datalog-encoded symbolic execution. The relationship to \sys is
sibling, not ancestral: Formulog puts SMT \emph{inside} the rule
body, and \sys runs SMT as a \emph{post-pass} over Souffl\'e's
findings. The two trade differently. Intra-rule SMT can prune
derivations as they form, but it needs a bespoke evaluation
strategy. Post-pass SMT keeps Souffl\'e at full relational
throughput and lets Z3 calls parallelise trivially over findings
(\S\ref{sec:methodology}). Concretely, our Datalog phase runs in
$\approx 2$ seconds on a 200K-fact libxml2 input and our SMT phase
in $\approx 1$ second on roughly 300 findings. Formulog's
benchmarks are in the same ballpark but on different workloads, so
we do not benchmark them head-to-head.

\paragraph{Static analysis narrows SMT (\sys's analytical lineage).}
The shape of \sys's symbex stage --- cheap static analysis
produces dataflow chains, SMT checks each chain --- is
well-established prior art. Saturn~\cite{xie2007saturn-toplas,
xie2005saturn-cav} introduced compositional SAT-based bug finding
using function summaries. Snugglebug~\cite{chandra2009snugglebug}
computes weakest preconditions backward from a target site,
demand-driven. Pinpoint~\cite{shi2018pinpoint}, built on
SVF~\cite{sui2016svf}, scales path-sensitive value-flow plus SMT to
million-line targets. \sys is best read as a re-application of this
lineage with three substantive changes. The dataflow substrate is
Datalog rather than a bespoke value-flow IR, so the analyst can
extend and audit it with the same idiom they would use for a new
taint rule. The substrate is fed by an LLM rather than a compiler,
so no build is required. And the SMT artifact is the SAT model
itself, not a yes-or-no verdict --- the model flows into \sys's
crash-synthesis stage. None of the four cited systems do crash
synthesis from the SMT witness.

\paragraph{LLMs for vulnerability detection.}
IRIS~\cite{li2024iris} demonstrated that LLMs are good at inferring
missing taint specifications --- sources, sinks, sanitisers --- and
that combining them with CodeQL improves recall on Java
vulnerabilities. \sys generalises the same observation. Once the
LLM has been told which schema of facts to emit, it is good at
producing those facts function by function, including the
structural facts (\texttt{ArithOp}, \texttt{Cast},
\texttt{FieldRead}, \texttt{Guard}) that go beyond IRIS's
specification-only role. Where IRIS uses the LLM to enrich a
classical analyser's input, \sys uses it to \emph{produce} that
input, replacing CodeQL's IR-lifting compiler rather than
augmenting it.

\paragraph{LLM-driven concolic and structured input generation.}
COTTONTAIL~\cite{tu2026cottontail} couples LLM-driven concolic
execution with built-target instrumentation for highly structured
input formats. Its LLM ``Constraint Mask / Flexible Mask''
decomposition --- the LLM marks the regions of the input
constrained by the path condition versus the regions the solver may
freely choose --- is the closest analogue to \sys's $P_c$ / $V_c$
scaffold split (\S\ref{sec:methodology}). The integration is again
sibling rather than ancestral. COTTONTAIL runs concolic on a built
target and the LLM steers input mutation, whereas \sys's synthesis
stage runs from a static SAT model with no concolic execution, and
the LLM is told both the scaffold (a known-good prefix to leave
undisturbed) and the $V_c$ bug condition. The two systems are
complementary: a static \sys finding could feed COTTONTAIL's
mutation budget on codebases where the target compiles.

\paragraph{Failure sketching and likely invariants.}
Kasikci~\textit{et al.}'s Gist~\cite{kasikci2015failure} introduced
adaptive backward-slice tracking with hardware watchpoints for
root-causing production failures. Its $\sigma$-doubling
slice-expansion idea is the source of \sys's Phase~E4
adaptive-slice heuristic. Sahoo~\textit{et al.}~\cite{sahoo2013likely}
introduced range invariants from a small number of normal-input
runs as a fault-localisation aid, and \sys repurposes them as a
precision filter on the SMT pass and adopts their
dependence-clustering idea (E1) directly. We deliberately do not
adopt their ddmin-style character rewriting for input generation,
because Phase~C uses an LLM with a file-format prior instead.

\paragraph{Symbolic execution and concolic platforms.}
KLEE~\cite{cadar2008klee} and SAGE-class whitebox
fuzzers~\cite{godefroid2008automated} are the established way to ask
SMT questions about whole-program behaviour, and
angr~\cite{shoshitaishvili2016sok} provides the canonical
binary-side platform for the same family of techniques. They
explore the program's path space directly. \sys's symbex stage
instead encodes only the def-use chain that a Datalog finding has
already identified, which is what makes its SMT cost
sub-second-per-finding rather than the open-ended exploration cost
KLEE-class tools incur. We see the two as different points on the
same axis: \sys gives up KLEE and angr's ability to discover paths
the analyst did not anticipate, in exchange for predictable
wall-clock cost at codebase scale.

\paragraph{Code-level fuzzing dataset.}
The Magma fuzzing benchmark~\cite{hazimeh2020magma} provides
ground-truth bug labels in libtiff, libxml2, libpng, and other
parsers. A larger version of the present evaluation would draw its
CVE set from Magma rather than from an ad-hoc list. We leave that
as future work.

%% file: sections/10_conclusion.tex
\section{Conclusion}
\label{sec:conclusion}

\sys is a small empirical argument for an old idea applied to a new
substrate. The old idea is that LLMs, declarative static analysis,
and SMT solving each have a part of the vulnerability-analysis task
they are uniquely good at, and that a pipeline assigning them
complementary roles is the natural arrangement. The new substrate
is source code without a build: a setting where an LLM is the only
tool that can lift the program to a usable representation, where a
Datalog rule mesh does that lifting's downstream work far more
cheaply than the LLM can on its own, and where an SMT post-pass can
then filter the results cheaply, because the rule mesh has already
shrunk the search.

The resulting pipeline re-discovers eight published CVE-class
issues end-to-end, two of them fuzzer-confirmed, and synthesises a
102-byte ASan-confirmed crashing input for one of them with no
fuzzer in the synthesis loop. An audit of libarchive HEAD surfaced
five more memory-safety bugs in current source, four of them
previously unreported, all filed upstream. A likely-invariant
filter collected from three corpus seeds eliminates 13.2\% of the
FFmpeg-demuxer feasible set, including the static false positive
the synthesiser had spent five attempts trying to crash. And the
cost-and-recall comparison with a frontier-model-only extractor is
one-sided: 90$\times$ faster, 70$\times$ cheaper, with no recall
loss on the bug-relevant functions.

The honest negatives matter, and we have presented them at the same
volume as the positives. The libxml2 source-side run missed sites
we expected to find until a bridging rule recovered them. The
FFmpeg demuxer slice leaves a 295-finding E2-filtered feasible set
we have not yet triaged by hand. And the most painful single
incident was a five-iteration synthesis attempt against what turned
out to be a static false positive that an earlier Phase~E2 run
would have caught.

We see \sys as a starting point. The task split it proposes --- an
LLM for fact extraction at the function level and crash synthesis
at the format level, Datalog for cross-function composition, SMT
for path feasibility, runtime invariants for sanity, and ASan as
the ground-truth oracle --- is meant as a template. We believe each
role can be tightened independently of the others, that a richer
rule mesh and a Magma-class evaluation~\cite{hazimeh2020magma}
would extend the empirical envelope, and that the compile-free
property is the right constraint to keep: it is what makes the
system feel, to an analyst opening an unfamiliar codebase, more
like \texttt{grep} than like CodeQL.

%% file: sections/A1_rules_appendix.tex
\appendix

\section{The \sys rule mesh}
\label{app:rules}

This appendix lists every output relation the rule mesh produces,
grouped by the bug family it targets. The relations are the
shared interface between Datalog (which materialises them) and
the symbex / synthesis stages (which consume them). Each relation
is keyed on \texttt{(func, addr, var)} or a small extension of
that tuple. The full schemas are in the public artifact's
\texttt{rules/} directory.

\paragraph{Taint propagation (interprocedural).}
\texttt{TaintSourceFunc} and \texttt{DangerousSink} are the two
catalogs the analyst can edit, and \texttt{TaintedVar} is the
fixpoint over them.

\begin{itemize}[leftmargin=*,itemsep=2pt]
\item \texttt{TaintedVar(func, var, addr, origin, kind)}
  --- variable \texttt{var} at \texttt{addr} carries a value
  derived from external input identified by \texttt{origin}.
\item \texttt{TaintedSink(caller, callee, ca, idx, var, risk, reason)}
  --- a tainted \texttt{var} reaches argument
  \texttt{idx} of a dangerous sink call.
\item \texttt{UnguardedTaintedSink(...)} --- as
  \texttt{TaintedSink} but with no CFG-reachable guard between
  the taint and the sink.
\end{itemize}

\paragraph{Integer / type-safety bugs.}

\begin{itemize}[leftmargin=*,itemsep=2pt]
\item \texttt{NarrowArithAtSink} --- arithmetic in a
  declared-narrow type whose result feeds a sink argument
  (e.g.\ a 32-bit add feeding a 64-bit \texttt{size\_t} parameter
  of \texttt{malloc}).
\item \texttt{TaintedNarrowArith} --- as above, with
  \texttt{TaintedVar} on the arithmetic destination.
\item \texttt{ImplicitTruncation} --- assignment of a wider type
  into a narrower lvalue without an explicit cast on a value that
  may not fit.
\item \texttt{TaintedTruncation} --- as above, with
  \texttt{TaintedVar} on the truncated value.
\item \texttt{SignedArgAtSink} --- a signed-typed value reaches a
  sink argument whose contract requires unsigned (e.g.\ a signed
  \texttt{int} feeding the \texttt{n} parameter of \texttt{memcpy}).
\item \texttt{TaintedSignedAtSink} --- signed-arg-at-sink with
  taint.
\item \texttt{SignednessMismatch} / \texttt{SignConfusionCast} ---
  cast between signed and unsigned of differing widths whose
  effective range silently changes.
\item \texttt{PotentialArithOverflow} --- arithmetic whose result
  type cannot represent the sum or product of the operand range.
\item \texttt{OverflowAtSink} --- \texttt{PotentialArithOverflow}
  whose result reaches a sink.
\item \texttt{TaintedOverflowAtSink} --- ditto with taint.
\end{itemize}

\paragraph{Counter / loop-bound bugs.}

\begin{itemize}[leftmargin=*,itemsep=2pt]
\item \texttt{UnboundedCounter} --- a variable incremented in a
  loop with no upper-bound guard.
\item \texttt{TaintedUnboundedCounter} --- as above, with taint.
\item \texttt{CounterUsedAsIndex} --- a loop counter used as a
  buffer index without a per-iteration bound check.
\item \texttt{TaintedCounterAsIndex} --- ditto with taint.
\item \texttt{TaintedLoopBound} --- the loop's terminating
  condition is itself a tainted variable.
\item \texttt{BufferOverflowInLoop} --- a tainted loop bound
  combined with a buffer write in the loop body and no
  per-iteration bound check.
\end{itemize}

\paragraph{Memory-safety bugs.}

\begin{itemize}[leftmargin=*,itemsep=2pt]
\item \texttt{UncheckedAlloc} --- the return value of an allocator
  is dereferenced without a NULL check on the CFG-reachable path.
\item \texttt{DoubleFree} --- two \texttt{free} calls on the same
  pointer without an intervening reassignment, with both calls
  reachable from one CFG entry.
\item \texttt{TaintedHeapObject} --- a heap object whose
  initialisation depends on taint (relevant for downstream
  type-confusion checks).
\item \texttt{AliasTaintedVar} --- alias-aware extension of
  \texttt{TaintedVar} for pointer-aliased writes.
\end{itemize}

\paragraph{Compound / summary relations.}

\begin{itemize}[leftmargin=*,itemsep=2pt]
\item \texttt{TypeSafetyFinding(func, addr, category, var, detail)}
  --- the union of all type-safety relations into one
  per-finding row, used by the symbex stage.
\item \texttt{MemSafetyFinding(...)} --- the analogous union for
  memory-safety relations.
\item \texttt{TaintReachableFunc} --- functions reachable from
  any \texttt{TaintSourceFunc} call site, used both for
  call-graph pruning and as a sanity check.
\end{itemize}

The complete rule mesh is approximately 30 rules across the five
files (\texttt{source\_taint.dl}, \texttt{source\_type\_safety.dl},
\texttt{source\_memsafety.dl}, \texttt{source\_interproc.dl},
\texttt{source\_sink\_pass.dl}); none of them is novel as a
declarative pattern --- the contribution of \sys is that the
facts that feed them come from an LLM rather than a compiler IR.

%% file: refs.bib
@inproceedings{jordan2016souffle,
  author    = {Herbert Jordan and Bernhard Scholz and Pavle Suboti{\'c}},
  title     = {Souffl{\'e}: On Synthesis of Program Analyzers},
  booktitle = {Computer Aided Verification (CAV)},
  year      = {2016},
  note      = {Project home: \url{https://souffle-lang.github.io/}}
}

@inproceedings{bravenboer2009doop,
  author    = {Martin Bravenboer and Yannis Smaragdakis},
  title     = {Strictly Declarative Specification of Sophisticated
               Points-to Analyses},
  booktitle = {Object-Oriented Programming, Systems, Languages,
               and Applications (OOPSLA)},
  year      = {2009}
}

@inproceedings{whaley2004bddbddb,
  author    = {John Whaley and Monica S. Lam},
  title     = {Cloning-based Context-sensitive Pointer Alias
               Analysis Using Binary Decision Diagrams},
  booktitle = {Programming Language Design and Implementation (PLDI)},
  year      = {2004}
}

@inproceedings{reps1995ifds,
  author    = {Thomas Reps and Susan Horwitz and Mooly Sagiv},
  title     = {Precise Interprocedural Dataflow Analysis via Graph
               Reachability},
  booktitle = {Principles of Programming Languages (POPL)},
  year      = {1995}
}

@inproceedings{andersen1994phd,
  author    = {Lars Ole Andersen},
  title     = {Program Analysis and Specialization for the {C}
               Programming Language},
  booktitle = {Ph.D.\ Thesis, DIKU, University of Copenhagen},
  year      = {1994}
}

@inproceedings{avgustinov2016ql,
  author    = {Pavel Avgustinov and Oege de Moor and Michael Peyton
               Jones and Max Sch{\"a}fer},
  title     = {{QL}: Object-oriented Queries on Relational Data},
  booktitle = {European Conference on Object-Oriented Programming (ECOOP)},
  year      = {2016}
}

@inproceedings{yamaguchi2014cpg,
  author    = {Fabian Yamaguchi and Nico Golde and Daniel Arp and
               Konrad Rieck},
  title     = {Modeling and Discovering Vulnerabilities with Code
               Property Graphs},
  booktitle = {{IEEE} Symposium on Security and Privacy (S\&P)},
  year      = {2014}
}

@misc{joern,
  author    = {{ShiftLeft, Inc.\ and the Joern community}},
  title     = {{Joern}: Open-source Code Analysis Platform for
               {C}/{C++}, Java, JavaScript, and others},
  howpublished = {\url{https://joern.io/}},
  year      = {2024},
  note      = {C/C++ frontend (\texttt{c2cpg}) parses source
               directly; \texttt{gcc} is used only optionally for
               system-header auto-discovery.}
}

@article{risse2025topscore,
  author    = {Niklas Risse and Jing Liu and Marcel B{\"o}hme},
  title     = {Top Score on the Wrong Exam: On Benchmarking in
               Machine Learning for Vulnerability Detection},
  journal   = {Proc.\ ACM Softw.\ Eng.},
  volume    = {2},
  number    = {ISSTA},
  pages     = {388--410},
  year      = {2025},
  doi       = {10.1145/3728887},
  note      = {arXiv:2408.12986}
}

@inproceedings{demoura2008z3,
  author    = {Leonardo de Moura and Nikolaj Bj{\o}rner},
  title     = {{Z3}: An Efficient {SMT} Solver},
  booktitle = {Tools and Algorithms for the Construction and
               Analysis of Systems (TACAS)},
  year      = {2008}
}

@inproceedings{cadar2008klee,
  author    = {Cristian Cadar and Daniel Dunbar and Dawson R. Engler},
  title     = {{KLEE}: Unassisted and Automatic Generation of
               High-coverage Tests for Complex Systems Programs},
  booktitle = {Operating Systems Design and Implementation (OSDI)},
  year      = {2008}
}

@inproceedings{godefroid2008automated,
  author    = {Patrice Godefroid and Michael Y. Levin and David
               Molnar},
  title     = {Automated Whitebox Fuzz Testing},
  booktitle = {Network and Distributed System Security (NDSS)},
  year      = {2008}
}

@inproceedings{shoshitaishvili2016sok,
  author    = {Yan Shoshitaishvili and Ruoyu Wang and Christopher
               Salls and Nick Stephens and Mario Polino and Andrew
               Dutcher and John Grosen and Siji Feng and Christophe
               Hauser and Christopher Kruegel and Giovanni Vigna},
  title     = {{SoK}: (State of) The Art of War: Offensive
               Techniques in Binary Analysis},
  booktitle = {{IEEE} Symposium on Security and Privacy (S\&P)},
  year      = {2016}
}

@misc{libfuzzer,
  author    = {{LLVM Project}},
  title     = {{libFuzzer}: A Library for Coverage-guided Fuzz Testing},
  howpublished = {\url{https://llvm.org/docs/LibFuzzer.html}},
  year      = {2024}
}

@misc{asan,
  author    = {Konstantin Serebryany and Derek Bruening and
               Alexander Potapenko and Dmitriy Vyukov},
  title     = {{AddressSanitizer}: A Fast Address Sanity Checker},
  howpublished = {\url{https://www.usenix.org/conference/atc12/technical-sessions/presentation/serebryany}},
  booktitle = {USENIX Annual Technical Conference (ATC)},
  year      = {2012}
}

@inproceedings{hazimeh2020magma,
  author    = {Ahmad Hazimeh and Adrian Herrera and Mathias Payer},
  title     = {{Magma}: A Ground-Truth Fuzzing Benchmark},
  booktitle = {Proceedings of the ACM on Measurement and Analysis
               of Computing Systems (POMACS)},
  year      = {2020},
  doi       = {10.1145/3428334}
}

@misc{li2024iris,
  author    = {Ziyang Li and Saikat Dutta and Mayur Naik},
  title     = {{IRIS}: {LLM}-Assisted Static Analysis for Detecting
               Security Vulnerabilities},
  howpublished = {arXiv:2405.17238},
  year      = {2024},
  note      = {\url{https://arxiv.org/abs/2405.17238}}
}

@inproceedings{sahoo2013likely,
  author    = {Swarup Kumar Sahoo and John Criswell and Chase Geigle
               and Vikram Adve},
  title     = {Using Likely Invariants for Automated Software Fault
               Localization},
  booktitle = {Architectural Support for Programming Languages and
               Operating Systems (ASPLOS)},
  year      = {2013}
}

@article{ernst2007daikon,
  author    = {Michael D. Ernst and Jeff H. Perkins and Philip J.
               Guo and Stephen McCamant and Carlos Pacheco and
               Matthew S. Tschantz and Chen Xiao},
  title     = {The {Daikon} System for Dynamic Detection of Likely
               Invariants},
  journal   = {Science of Computer Programming},
  volume    = {69},
  number    = {1--3},
  pages     = {35--45},
  year      = {2007}
}

@inproceedings{kasikci2015failure,
  author    = {Baris Kasikci and Benjamin Schubert and Cristiano
               Pereira and Gilles Pokam and George Candea},
  title     = {Failure Sketching: A Technique for Automated Root
               Cause Diagnosis of In-Production Failures},
  booktitle = {ACM Symposium on Operating Systems Principles (SOSP)},
  year      = {2015}
}

@misc{tree-sitter,
  author    = {Max Brunsfeld and others},
  title     = {Tree-sitter: An Incremental Parsing System for
               Programming Tools},
  howpublished = {\url{https://tree-sitter.github.io/tree-sitter/}},
  year      = {2024}
}

@inproceedings{tu2026cottontail,
  author    = {Haoxin Tu and Seongmin Lee and Yuxian Li and
               Peng Chen and Lingxiao Jiang and Marcel B{\"o}hme},
  title     = {{Cottontail}: Large Language Model-Driven Concolic
               Execution for Highly Structured Test Input
               Generation},
  booktitle = {{IEEE} Symposium on Security and Privacy (S\&P)},
  year      = {2026}
}

@article{bembenek2020formulog,
  author    = {Aaron Bembenek and Michael Greenberg and Stephen Chong},
  title     = {{Formulog}: Datalog for {SMT}-Based Static Analysis},
  journal   = {Proceedings of the ACM on Programming Languages
               (PACMPL), {OOPSLA}},
  volume    = {4},
  number    = {OOPSLA},
  year      = {2020},
  note      = {arXiv:2009.08361}
}

@article{bembenek2024fast,
  author    = {Aaron Bembenek and Michael Greenberg and Stephen Chong},
  title     = {Making {Formulog} Fast: An Argument for Unconventional
               {Datalog} Evaluation},
  journal   = {Proceedings of the ACM on Programming Languages
               (PACMPL), {OOPSLA2}},
  volume    = {8},
  number    = {OOPSLA2},
  year      = {2024},
  note      = {arXiv:2408.14017}
}

@inproceedings{xie2005saturn-cav,
  author    = {Yichen Xie and Alex Aiken},
  title     = {{Saturn}: A {SAT}-Based Tool for Bug Detection},
  booktitle = {Computer Aided Verification (CAV)},
  year      = {2005}
}

@article{xie2007saturn-toplas,
  author    = {Yichen Xie and Alex Aiken},
  title     = {{Saturn}: A Scalable Framework for Error Detection
               Using {Boolean} Satisfiability},
  journal   = {ACM Transactions on Programming Languages and Systems
               (TOPLAS)},
  volume    = {29},
  number    = {3},
  year      = {2007}
}

@inproceedings{chandra2009snugglebug,
  author    = {Satish Chandra and Stephen J. Fink and Manu Sridharan},
  title     = {{Snugglebug}: A Powerful Approach to Weakest
               Preconditions},
  booktitle = {Programming Language Design and Implementation (PLDI)},
  year      = {2009}
}

@inproceedings{shi2018pinpoint,
  author    = {Qingkai Shi and Xiao Xiao and Rongxin Wu and Jinguo
               Zhou and Gang Fan and Charles Zhang},
  title     = {{Pinpoint}: Fast and Precise Sparse Value Flow
               Analysis for Million Lines of Code},
  booktitle = {Programming Language Design and Implementation (PLDI)},
  year      = {2018}
}

@inproceedings{sui2016svf,
  author    = {Yulei Sui and Jingling Xue},
  title     = {{SVF}: Interprocedural Static Value-Flow Analysis in
               {LLVM}},
  booktitle = {International Conference on Compiler Construction
               (CC)},
  year      = {2016}
}

@misc{cve-2023-45676,
  author    = {{GitHub Security Lab}},
  title     = {{CVE-2023-45676}: Integer overflow leading to
               heap-based buffer overflow in {stb\_vorbis}
               comment-parsing path ({GHSL-2023-166})},
  howpublished = {\url{https://nvd.nist.gov/vuln/detail/CVE-2023-45676}},
  year      = {2023}
}

@misc{cve-2023-45681,
  author    = {{GitHub Security Lab}},
  title     = {{CVE-2023-45681}: Integer overflow in {stb\_vorbis}
               comment-list allocation ({GHSL-2023-171})},
  howpublished = {\url{https://nvd.nist.gov/vuln/detail/CVE-2023-45681}},
  year      = {2023}
}

@misc{cve-2025-57052,
  author    = {{NVD}},
  title     = {{CVE-2025-57052}: Out-of-bounds access in {cJSON}
               JSON-Pointer handling},
  howpublished = {\url{https://nvd.nist.gov/vuln/detail/CVE-2025-57052}},
  year      = {2025}
}

@misc{cve-2023-53154,
  author    = {{NVD}},
  title     = {{CVE-2023-53154}: Heap buffer over-read in {cJSON}
               string parsing},
  howpublished = {\url{https://nvd.nist.gov/vuln/detail/CVE-2023-53154}},
  year      = {2023}
}

@misc{cve-2025-6021,
  author    = {{NVD}},
  title     = {{CVE-2025-6021}: Integer overflow in
               {libxml2} \texttt{xmlBuildQName}},
  howpublished = {\url{https://nvd.nist.gov/vuln/detail/CVE-2025-6021}},
  year      = {2025}
}

@misc{stb,
  author    = {Sean Barrett},
  title     = {{stb} single-file public-domain libraries for {C}/{C++}},
  howpublished = {\url{https://github.com/nothings/stb}},
  year      = {2024}
}

@misc{cjson,
  author    = {Dave Gamble and contributors},
  title     = {{cJSON}: An ultralightweight {JSON} parser in {ANSI} {C}},
  howpublished = {\url{https://github.com/DaveGamble/cJSON}},
  year      = {2024}
}

@misc{libxml2,
  author    = {Daniel Veillard and contributors},
  title     = {{libxml2}: The {XML} {C} parser and toolkit of {Gnome}},
  howpublished = {\url{https://gitlab.gnome.org/GNOME/libxml2}},
  year      = {2024}
}

@misc{ffmpeg,
  author    = {{The FFmpeg Project}},
  title     = {{FFmpeg}: A Complete, Cross-platform Solution to
               Record, Convert and Stream Audio and Video},
  howpublished = {\url{https://ffmpeg.org/}},
  year      = {2024}
}

@misc{cve-2023-38545,
  author    = {{NVD}},
  title     = {{CVE-2023-38545}: {SOCKS5} heap buffer overflow in
               {curl}},
  howpublished = {\url{https://nvd.nist.gov/vuln/detail/CVE-2023-38545}},
  year      = {2023}
}

@misc{cve-2025-10148,
  author    = {{NVD}},
  title     = {{CVE-2025-10148}: Predictable per-connection mask key
               in {curl} {WebSocket}},
  howpublished = {\url{https://nvd.nist.gov/vuln/detail/CVE-2025-10148}},
  year      = {2025}
}

@misc{libarchive,
  author    = {Tim Kientzle and contributors},
  title     = {{libarchive}: Multi-format archive and compression
               library},
  howpublished = {\url{https://github.com/libarchive/libarchive}},
  year      = {2026}
}

@misc{curl-socks-fix-8.4.0,
  author    = {{curl project}},
  title     = {{SOCKS4a} long-username/hostname heap overflow
               (\texttt{strcpy} joint-bound check), HackerOne
               disclosure},
  howpublished = {\url{https://github.com/curl/curl/commit/01057d6161}},
  year      = {2023},
  note      = {Patch \texttt{01057d6161}, curl 8.4.0}
}

@misc{libarchive-issue-xar-reader,
  author    = {{Anonymous}},
  title     = {{xar} reader: \texttt{file\_free()} misses
               \texttt{archive\_string\_free} for
               \texttt{fflags\_text} (heap leak)},
  howpublished = {\url{https://github.com/libarchive/libarchive/issues/3058}},
  year      = {2026},
  note      = {Filed 2026-05-21; acknowledged with fix
               PR \#3060 within $\sim$18 hours}
}

@misc{libarchive-issue-xar-writer,
  author    = {{Anonymous}},
  title     = {{xar} writer: \texttt{make\_fflags\_entry()} OOB read
               into \texttt{.rodata} past string-literal null
               terminator},
  howpublished = {\url{https://github.com/libarchive/libarchive/issues/3059}},
  year      = {2026},
  note      = {Filed 2026-05-21; concurrent with maintainer's
               in-flight fix PR \#3041
               (\texttt{GHSA-wfvr-54j8-47r9})}
}

@misc{libarchive-pr-3041,
  author    = {Tobias Stoeckmann},
  title     = {{xar}: Fix writer OOB accesses with fflags
               (resolves \texttt{GHSA-wfvr-54j8-47r9})},
  howpublished = {\url{https://github.com/libarchive/libarchive/pull/3041}},
  year      = {2026},
  note      = {Opened 2026-05-16; concurrent in-flight fix for the
               same XAR-writer OOB read site we independently
               identified two days later}
}

@misc{libarchive-issue-warc-writer,
  author    = {{Anonymous}},
  title     = {{warc} writer: \texttt{\_warc\_header()} leaks
               \texttt{archive\_string} on \texttt{\_popul\_ehdr}
               overflow},
  howpublished = {GitHub issue (URL pending: will be filled when filed)},
  year      = {2026}
}

@misc{libarchive-issue-7zip-writer,
  author    = {{Anonymous}},
  title     = {{7zip} writer: \texttt{file\_new()} leaks
               \texttt{file->utf16name} on symlink UTF-8
               conversion failure},
  howpublished = {GitHub issue (URL pending: will be filled when filed)},
  year      = {2026}
}

@misc{libarchive-issue-3053,
  author    = {{Anonymous}},
  title     = {{cpio} reader: \texttt{record\_hardlink()}
               {UAF}/double-free regression --- \texttt{free(le)}
               leaves dangling pointer in
               \texttt{cpio->links\_head}},
  howpublished = {\url{https://github.com/libarchive/libarchive/issues/3053}},
  year      = {2026},
  note      = {Filed 2026-05-20; acknowledged with fix PR \#3055 in
               7 hours, reporter credited as commit author}
}
